\documentclass[fleqn,usenatbib]{mnras}

\usepackage{newtxtext,newtxmath}

\usepackage{lineno} 

\usepackage[T1]{fontenc}

\DeclareRobustCommand{\VAN}[3]{#2}
\let\VANthebibliography\thebibliography
\def\thebibliography{\DeclareRobustCommand{\VAN}[3]{##3}\VANthebibliography}

\usepackage{graphicx}	
\usepackage{amsmath}	

\title[Indus Density Variation]{Epicyclic Density Variations in the Indus Stellar Stream}

\author[Y. Yang et al.]{Yong~Yang,$^{1}$\thanks{E-mail: yong.yang@sydney.edu.au}
    Geraint~F.~Lewis,$^{1}$
    Ting~S.~Li,$^{2,3,4}$
    Sarah~L.~Martell,$^{5}$
    Denis~Erkal,$^{6}$
    Alexander~P.~Ji,$^{7,8,9}$
    \newauthor
    Sergey~E.~Koposov,$^{10,11}$
    Daniel~B.~Zucker,$^{12,13}$
    Andrew~B.~Pace,$^{14}$\thanks{Galaxy Evolution and Cosmology (GECO) Fellow}
    Lara~R.~Cullinane,$^{15}$
    Gary~S.~Da~Costa,$^{16}$
    \newauthor
    Kyler~Kuehn,$^{17}$
    Guilherme~Limberg,$^{8}$
    Gustavo~E.~Medina,$^{2,3}$
    and $S^5$~Collaboration
\\
$^{1}$Sydney Institute for Astronomy, School of Physics, A28, The University of Sydney, NSW 2006, Australia\\
$^{2}$Department of Astronomy and Astrophysics, University of Toronto, 50 St. George Street, Toronto ON, M5S 3H4, Canada\\
$^{3}$Dunlap Institute for Astronomy \& Astrophysics, University of Toronto, 50 St George Street, Toronto, ON M5S 3H4, Canada\\
$^{4}$Data Sciences Institute, University of Toronto, 17th Floor, Ontario Power Building, 700 University Ave, Toronto, ON M5G 1Z5, Canada\\
$^{5}$School of Physics, University of New South Wales, Sydney, NSW 2052, Australia\\
$^{6}$Department of Physics, University of Surrey, Guildford GU2 7XH, UK\\
$^{7}$Department of Astronomy \& Astrophysics, University of Chicago, 5640 S Ellis Avenue, Chicago, IL 60637, USA\\
$^{8}$Kavli Institute for Cosmological Physics, University of Chicago, Chicago, IL 60637, USA\\
$^{9}$NSF-Simons AI Institute for the Sky (SkAI), 172 E. Chestnut St., Chicago, IL 60611, USA\\
$^{10}$Institute for Astronomy, University of Edinburgh, Royal Observatory, Blackford Hill, Edinburgh EH9 3HJ, UK\\
$^{11}$Institute of Astronomy, University of Cambridge, Madingley Road, Cambridge CB3 0HA, UK\\
$^{12}$School of Mathematical and Physical Sciences, Macquarie University, Sydney, NSW 2109, Australia\\
$^{13}$Macquarie University Research Centre for Astrophysics and Space Technologies, Sydney, NSW 2109, Australia\\
$^{14}$Department of Astronomy, University of Virginia, 530 McCormick Road, Charlottesville, VA 22904, USA\\
$^{15}$Leibniz-Institut f{\"u}r Astrophysik Potsdam (AIP), An der Sternwarte 16, D-14482 Potsdam, Germany\\
$^{16}$Research School of Astronomy and Astrophysics, Australian National University, Canberra, ACT 2611, Australia\\
$^{17}$Lowell Observatory, 1400 W Mars Hill Rd, Flagstaff,  AZ 86001, USA
}

\date{Accepted XXX. Received YYY; in original form ZZZ}

\pubyear{\the\year{}}

\begin{document}


\label{firstpage}
\pagerange{\pageref{firstpage}--\pageref{lastpage}}
\maketitle

\begin{abstract}
Longitudinal density fluctuations observed in stellar streams can result from gravitational interactions with massive perturbers in the Milky Way, such as dark matter subhalos. Analysing these density variations provides a powerful probe of properties (motion, mass, size, etc.) of the perturbing objects. However, caution is needed because density variations may arise naturally from internal dynamics of streams, namely epicycles. In this work, we focus on the Indus stellar stream, a remnant of an ancient dwarf satellite of the Galaxy. An Indus stream spanning $\sim 90\degr$ is revealed in the southern Galactic sky using a comprehensive matched-filter analysis utilizing data from the \textit{Gaia} mission. A spatial density model is fitted to the filtered map to quantitatively characterize the morphology, which demonstrates episodic density peaks and gaps in the stream. Through N-body simulations, we show that there are strong epicyclic motions of stars happening during tidal disruptions. The present-day longitudinal densities from simulations are comparable to the measurement from data, with similar numbers and locations of peaks and gaps, suggesting that the observed density should mainly be caused by epicycles. We also find that a cuspy dark matter halo for the Indus dwarf is likely to produce milder stellar epicyclic peaks compared to a cored halo which results in steeper peaks. This arises from different instantaneous mass loss due to distinct central mass distributions of halos, where a cored halo usually leads to severer tidal stripping. The observed density exhibits moderate peak sharpness, implying that Indus may have originally possessed a cuspy halo.
\end{abstract}

\begin{keywords}
Galaxy: halo -- Galaxy: structure -- Galaxy: kinematics and dynamics
\end{keywords}



\section{Introduction} \label{sec:intro}

In the Lambda Cold Dark Matter ($\Lambda$CDM) paradigm, the universe begins with small density fluctuations, where denser regions gravitationally attract matter. Subsequently, small dark matter halos are formed, which merge over time into larger structures, laying the foundation for star and galaxy formation \citep[e.g.][]{1999ApJ...524L..19M,1999ApJ...522...82K,2008MNRAS.391.1685S,2008Natur.454..735D}. Our host galaxy, the Milky Way, exhibits significant evidence to support such a hierarchical formation prediction; e.g., extant dwarf galaxies \citep[e,g,][]{2007ApJ...654..897B,2015ApJ...807...50B,2020MNRAS.495.2554E,2020ApJ...893..121P,2021ApJ...921...32J,2022ApJ...940..136P,2025OJAp....8E.112P,2025arXiv250902546S}, the ongoing Sagittarius merger \citep{1994Natur.370..194I,2010ApJ...714..229L,2019ApJ...886..154Y,2023ApJ...946...66L}, spatially-visible debris of previously disrupted dwarfs \citep{2018ApJ...862..114S,2022ApJ...930..103Y,2023MNRAS.521.4936K}, and more ancient accretions identified using integrals of motion \citep{2018Natur.563...85H,2020ApJ...901...48N,2024MNRAS.527.9767L,2025ApJ...983...65Y} or chemical abundances \citep{2020AJ....160..181J,2021SCPMA..6439562Z,2024ApJ...966..174Z,2024MNRAS.530.2512L}.

However, not all small halos merge completely, and many survive as subhalos orbiting within the larger halo \citep[e.g.][]{2007ApJ...654..897B,2008Natur.454..735D,2015ApJ...807...50B,2020MNRAS.495.2554E,2020ApJ...893..121P,2021ApJ...921...32J,2022ApJ...940..136P,2025OJAp....8E.112P}. As a result, the Milky Way is expected to be embedded within a vast population of leftover dark matter subhalos, many of which are too small ($\la 10^8$ M$_{\sun}$) to host visible stars. Thus, they remain undetected through direct observations \citep{2015ARA&A..53...51S}. Validating the existence of these dark substructures is crucial for testing the $\Lambda$CDM framework. Two promising observational approaches have emerged to probe these elusive subhalos. The first is gravitational lensing, which can reveal their presence through distortions in the light from background sources \citep[e.g.][]{2012Natur.481..341V,2016ApJ...823...37H,2024MNRAS.528.7564B}. The second is stellar streams, whose delicate structures can be perturbed by interactions with dark matter clumps. This offers a dynamic tracer of the invisible architecture of our Galaxy \citep[e.g.][]{2002MNRAS.332..915I,2002ApJ...570..656J,2012ApJ...748...20C,2015MNRAS.450.1136E,2015MNRAS.454.3542E,2016MNRAS.463..102E,2019ApJ...880...38B,2021MNRAS.502.2364B,2024arXiv240402953H}.

A stellar stream is made of stars stripped from a globular cluster or dwarf galaxy \citep[see][]{2025NewAR.10001713B}. These stars usually follow very similar orbits to that of the progenitor since orbital energy does not change significantly after they become unbound to their host \citep{2013MNRAS.433.1813S,2021ApJ...914..123I,2025MNRAS.539.2718P}, producing long, thin streams that are sensitive to perturbations from other massive bodies. As a result, streams have been considered valuable tools to probe dark matter subhalos \citep[e.g.][]{2002MNRAS.332..915I,2002ApJ...570..656J}. For example, stellar counts may vary along a stream to form some underdensity (gap) and overdensity (peak) features due to flybys of subhalos \citep{2015MNRAS.450.1136E}. However, effects from baryonic structures in the Galaxy make this question more complicated because just like dark subhalos, there are chances that streams are influenced by the central bar \citep{2017MNRAS.470...60E,2017NatAs...1..633P}, spiral arms \citep{2019MNRAS.484.2009B}, giant molecular clouds \citep{2016MNRAS.463L..17A}, globular clusters \citep{2025A&A...699A.289F}, or dwarf galaxies \citep{2023A&A...669A.102W}. Besides external shocks, density clumps could also arise naturally from internal dynamical mechanisms, such as full dissolution of stream progenitors \citep{2019MNRAS.485.5929W,2025MNRAS.538..454R}, and epicyclic motions of stars in streams \citep{2008MNRAS.387.1248K,2009MNRAS.392..969J,2010MNRAS.401..105K,2012A&A...546L...7M,2020ApJ...891..161I}. Given the complexity of multiple contributing factors to density variations, accurately localizing and quantifying dark matter subhalos using stellar streams remains a significant observational and modelling challenge.

The Indus stellar stream, one of $>$ 100 known Galactic streams so far \citep{2023MNRAS.520.5225M}, was first discovered using deep photometric data from the Dark Energy Survey \citep[DES;][]{2018ApJ...862..114S} and further observed by the Southern Stellar Stream Spectroscopic Survey \citep[$S^5$;][]{2019MNRAS.490.3508L}. Indus is believed to be a remnant of a dwarf galaxy based on its kinematics and chemistry, including the stream width \citep{2018ApJ...862..114S}, velocity and metallicity dispersion \citep{2022ApJ...928...30L}, as well as abundances \citep{2020AJ....160..181J,2021ApJ...915..103H}.

Compared to previous studies mostly focusing on densities of dynamically cold streams with globular cluster origins, here we explore the density variation of Indus which has a dwarf galaxy origin. This paper is organized as follows. Section~\ref{sec:data} briefly introduces the data sets employed in this study. Section~\ref{sec:extended} highlights the detection of Indus stream. Section~\ref{sec:density} characterizes the stream's spatial morphology. Section~\ref{sec:model} details N-body modelling of the stream. Section~\ref{sec:future} lays out future prospects. Section~\ref{sec:summary} presents conclusions of the work.

\section{Data} \label{sec:data}

We conduct this work based on the third \textit{Gaia} data release \cite[\textit{Gaia} DR3;][]{2023A&A...674A...1G}, which provides us with high-quality astrometry and photometry. The parallax zeropoint is corrected using codes from \citet{2021A&A...649A...4L}. In order to ensure good astrometric and photometric measurements, those stars satisfying \texttt{ruwe} $<$ 1.4 and $|C^*|$ $<$ 3$\sigma_{C^*}$ are retained, where \texttt{ruwe} is the renormalized unit weight error \citep{2021A&A...649A...2L}, $C^*$ is the corrected BP and RP flux excess, and $\sigma_{C^*}$ is a fitted scatter of $C^*$ as a function of $G$ magnitude \citep[see Section~9.4 in][]{2021A&A...649A...3R}. The photometry is extinction-corrected using the dust map from \citet{1998ApJ...500..525S} and \citet{2011ApJ...737..103S}, assuming $R_V$ = 3.1, $A_V/E(B-V)_\mathrm{SFD}$ = 2.742, $A_G/A_V$ = 0.83627, $A_{BP}/A_V$ = 1.08337, and $A_{RP}/A_V$ = 0.63439.\footnote{These extinction ratios are taken from the Padova model site, \url{https://stev.oapd.inaf.it/cgi-bin/cmd}.} We further calculate magnitude errors in the BP and RP bands using a propagation of the flux errors\footnote{\url{https://vizier.cds.unistra.fr/viz-bin/VizieR-n?-source=METAnot&catid=1350&notid=63&-out=text}.} and obtain the colour error $\sqrt{\sigma^2_{BP}+\sigma^2_{RP}}$. Since astrometric and photometric uncertainties become larger for fainter stars, we only use stars with $G_0 < 20$ mag for Indus stream identification and modelling.

In addition to the proper motions from \textit{Gaia}, we also have radial velocities and metallicities from the $S^5$ survey \citep{2019MNRAS.490.3508L}, which is aimed at obtaining deep insights into known streams in the southern hemisphere. Observations were initiated in 2018 with the Two-degree Field (2dF) fibre positioner \citep{2002MNRAS.333..279L} coupled with the dual-arm AAOmega spectrograph \citep{2006SPIE.6269E..0GS} on the 3.9-m Anglo-Australian Telescope (AAT). The $S^5$ observations provide spectroscopic measurements for stream candidate members including radial velocity and metallicity determined by the \texttt{rvspecfit} pipeline \citep{2019ascl.soft07013K}, as well as astrometry crossmatched from \textit{Gaia} DR3. Here we employ $S^5$ data, specifically published member stars belonging to the Indus stream in \citet{2022ApJ...928...30L}. 

\section{Indus detection} \label{sec:extended}

An extended Indus has previously been implied in \citet{2018ApJ...862..114S}, where the stream was $\sim 20\degr$ long and truncated due to data coverage. \citet{2024ApJ...967...89I} further detected three discrete Indus segments spanning $\sim 80\degr$. Here we conduct a detailed matched filter search to present a full view of the Indus stream in southern Galactic sky.

\subsection{Matched Filters}

Prior to the \textit{Gaia} era, the detection of stellar streams primarily relied on the application of matched filters \citep{2002AJ....124..349R} in a colour-magnitude diagram (CMD), under the assumption that streams' stellar populations follow specific isochrones \citep[e.g.][]{2006ApJ...643L..17G,2014MNRAS.442L..85K}. The arrival of \textit{Gaia} makes it feasible to further incorporate proper motion filters to enhance the contrast between streams and foreground stars \citep[e.g.][]{2019ApJ...884..174G,2023ApJ...953..130Y,2025ApJ...995...15C}. Here we adopt this approach to filter our \textit{Gaia} data set in an ``unweighted'' way, that is, to directly select or reject stars, so that star counts will follow Poisson statistics \citep{2018ApJ...862..114S,2022AJ....163...18F}.

\subsubsection{Kinematic Filters} \label{subsec:k-filter}

We build kinematic filters in on-sky positions and proper motions by referring to a trial Indus model.

To perform the required simulations, we employ a Galactic potential comprising the Milky Way and the Large Magellanic Cloud (LMC). The Milky Way is the axisymmetrised version from \citet{2024A&A...692A.216H}, while the LMC is represented by a truncated NFW profile \citep{1997ApJ...490..493N} with a mass of $1.5\times10^{11}$ M$_{\sun}$ \citep{2019MNRAS.487.2685E,2021ApJ...923..149S,2023MNRAS.521.4936K} and a scale radius of 10.84 kpc. The time-dependent potential also takes into account the LMC's gravitational pull on the Milky Way which causes the latter to be a non-inertial reference frame. All of the LMC-related parameters are taken from \citet{2021MNRAS.501.2279V}. We refer to this potential model as \texttt{potMWLMC} later. To transform between ICRS and Galactocentric frames, we adopt attributes of ``v4.0'' in \texttt{astropy} \citep{2022ApJ...935..167A}, that is, the Sun's position ($R_{\odot}$, $z_{\odot}$) = (8.122, 0.0208) kpc \citep{2018A&A...615L..15G,2019MNRAS.482.1417B} and velocities ($V_{R,\odot}$, $V_{\phi,\odot}$, $V_{z,\odot}$) = ($-$12.9, 245.6, 7.78) km~s$^{-1}$ \citep{2018RNAAS...2..210D}.

The stream orbit is fitted using Indus members from $S^5$ \citep{2022ApJ...928...30L}. There are 75 stars having full six-dimension information with distances derived from a fitted relation (Table~1 therein). The progenitor's current location is assumed to be at Galactic longitude $l=333\degr$, roughly equal to the midpoint of these stars. The other five coordinates of the progenitor (Galactic latitude $b$, distance $d_{\sun}$, proper motions $\mu^*_{\alpha}$ and $\mu_{\delta}$, radial velocity $V_r$) are set free to be sampled with the Markov Chain Monte Carlo (MCMC) sampler \texttt{emcee} \citep{2013PASP..125..306F}. For each sample of the progenitor coordinates, we integrate the corresponding orbit using \texttt{agama} \citep{2019MNRAS.482.1525V} and evaluate its consistency with measurements of 75 $S^5$ stars in the above five dimensions (here $l$ is taken as the stream longitude) by assuming a Gaussian likelihood and a uniform prior. The best-fit results are ($l$, $b$, $d_{\sun}$, $\mu^*_{\alpha}$, $\mu_{\delta}$, $V_r$) $=$ ($333\degr$, $-49.13^{+0.09}_{-0.08}\degr$, $14.90\pm0.04$ kpc, $3.22\pm0.01$ mas~yr$^{-1}$, $-5.15\pm0.01$ mas~yr$^{-1}$, $-55.07^{+0.17}_{-0.18}$ km~s$^{-1}$).

\begin{figure}
    \centering
    \includegraphics[width=\columnwidth]{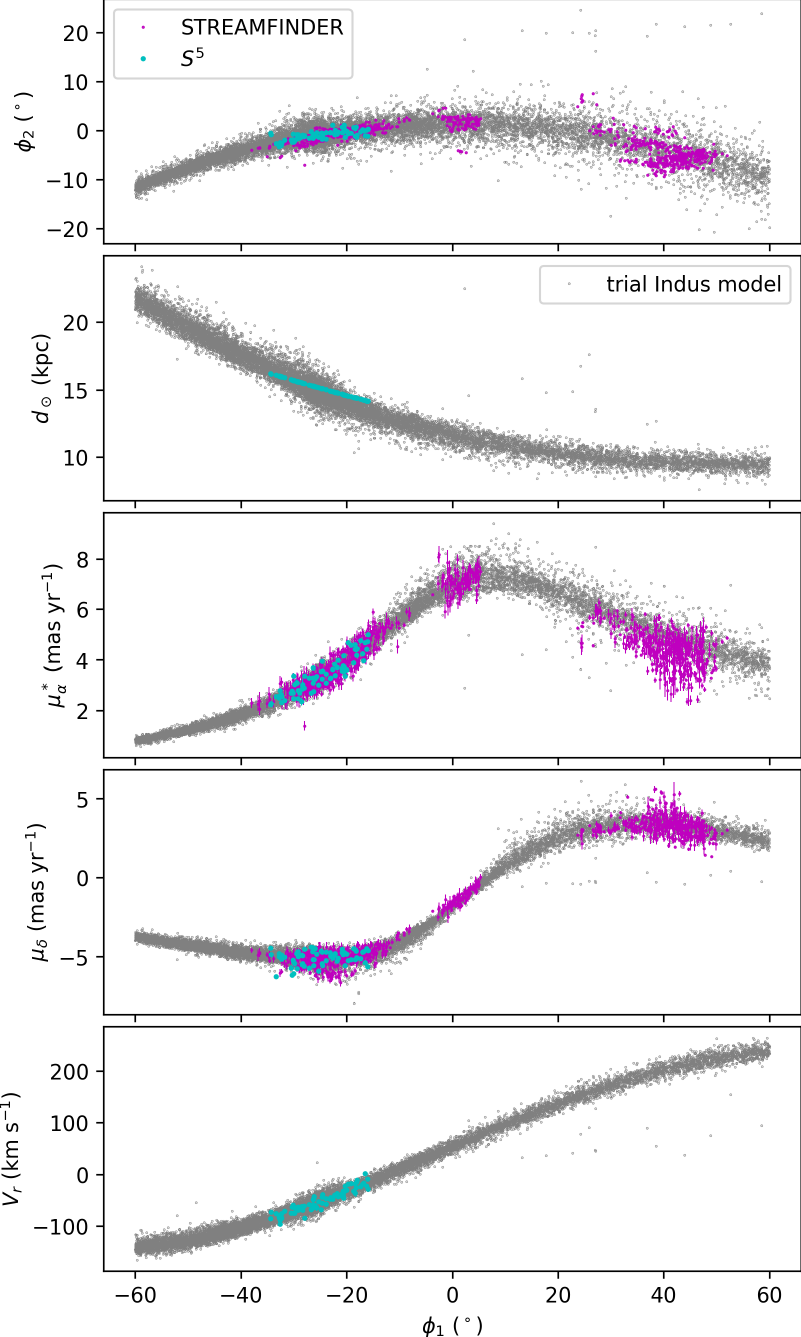}
    \caption{Comparisons between the trial model and real data of the Indus stream in phase spaces. Five panels display $\phi_2$, $d_\odot$, $\mu^*_{\alpha}$, $\mu_{\delta}$, and $V_r$, respectively, as a function of $\phi_1$. Grey dots show the model, while cyan and magenta points show Indus members from $S^5$ \citep{2022ApJ...928...30L} and \texttt{STREAMFINDER} \citep{2024ApJ...967...89I}.}
    \label{fig:modelvsdata}
\end{figure}

A trial model stream is produced with the particle-spray technique from \citet{2019MNRAS.487.2685E}. Briefly, the best-fit orbit is rewound to 3 Gyr before present, with a time spacing of 1 Myr: thus 3000 steps in total. At each timestep from the beginning, particles are released at Lagrange points with assigned random velocities drawn from a dispersion = 7 km~s$^{-1}$, matching the measured dispersion of the stream \citep{2022ApJ...928...30L}. These particles are evolved in the above \texttt{potMWLMC} potential in the presence of the Indus progenitor potential as well, which is modelled by a NFW profile with a scale radius of 0.1 kpc and a scale mass of $6 \times 10^6$ M$_{\sun}$ \citep{2018ApJ...862..114S} where the mass will linearly decrease over time. See e.g. \citet{2025ApJ...984..189Y} for more details about the spray method. Figure~\ref{fig:modelvsdata} presents a direct comparison between the trial model and real data of Indus in phase spaces. Here a stream-aligned coordinate system ($\phi_1$, $\phi_2$) is defined by two endpoints (RA, Dec) = ($340\degr$, $-60\degr$) and ($80\degr$, $-60\degr$) using \texttt{gala} package \citep{2017JOSS....2..388P}. The corresponding rotation matrix is given by 
\begin{equation}
R = 
\begin{bmatrix}
 0.3013136404 & 0.1739635114 & -0.9375216194 \\
 -0.5000000000 & 0.8660254038 & 0.0000000000  \\
 0.8119175390 & 0.4687608097 & 0.3479270228
\end{bmatrix}
.
\end{equation}
In this coordinate, the progenitor is at ($\phi_1$, $\phi_2$) = ($-25.82\degr$, $-0.82\degr$). The grey dots show the model particles, while coloured points show the data including Indus members from $S^5$ \citep{2022ApJ...928...30L} and \texttt{STREAMFINDER} \citep{2024ApJ...967...89I}. The latter ones are only plotted in $\phi_2$, $\mu^*_\alpha$, and $\mu_\delta$ due to the lack of $d_\odot$ and $V_r$. In general, we can note excellent agreements between the model and data, either in the stream tracks or dispersions. There is only a little mismatch to the rightmost end in the $\mu^*_\alpha$ panel. We consider that this trial model has been good enough to serve as an estimate of the real Indus in the observed data. 

\begin{figure}
    \centering
    \includegraphics[width=0.9\columnwidth]{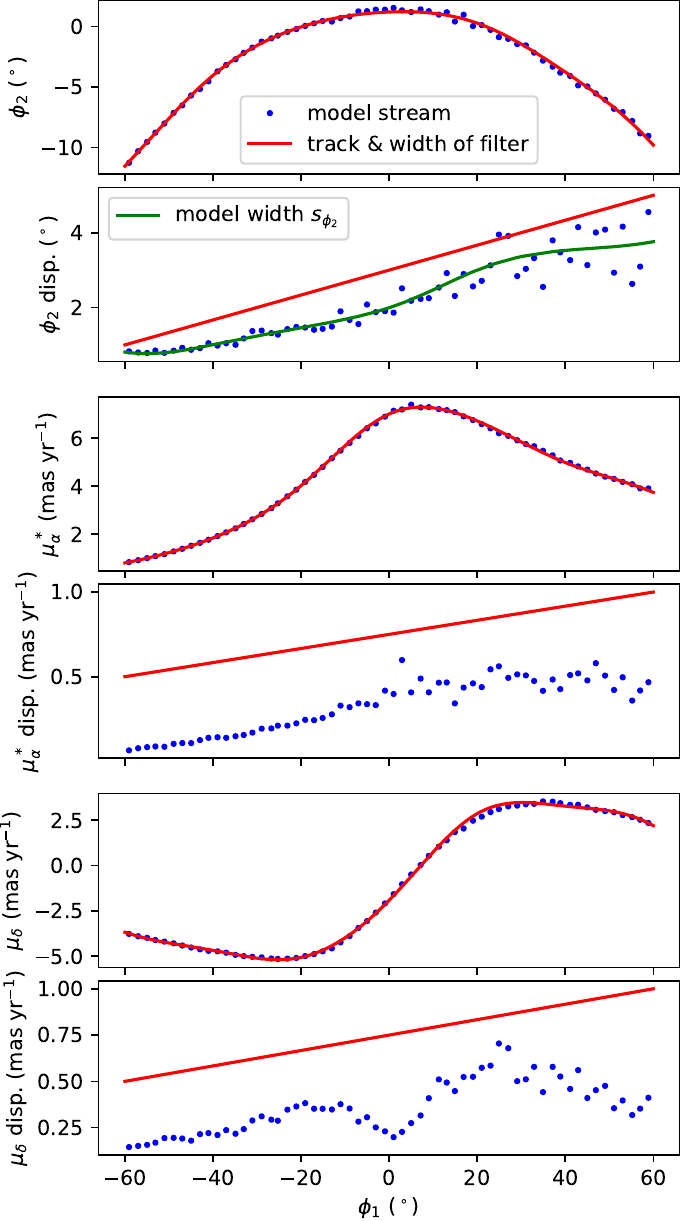}
    \includegraphics[width=0.9\columnwidth]{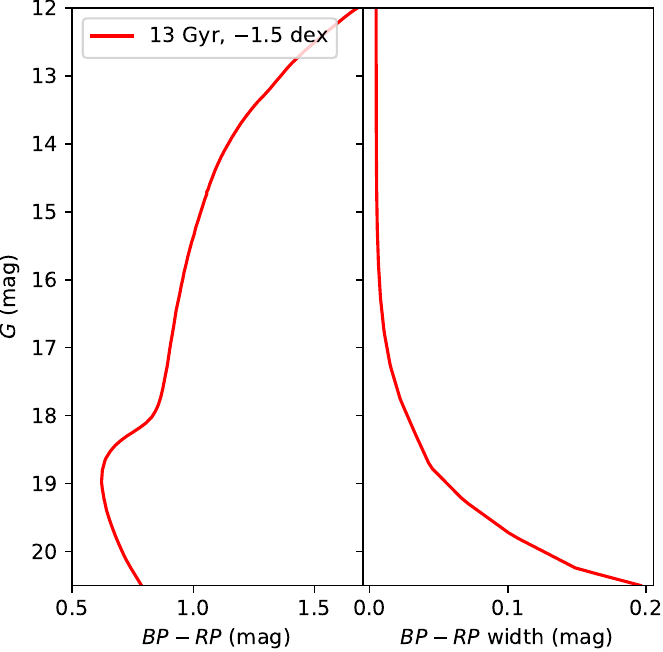}
    \caption{Top six panels: kinematic filters created based on the trial model of the Indus stream. The three pairs of panels show the mean track and width in $\phi_2$, $\mu^*_{\alpha}$, and $\mu_{\delta}$, respectively. Blue points are computed from the model stream, while red lines are tracks and widths designed for filtering. The green line is the model width $s_{\phi_2}$ fitted using splines. Bottom two panels: the CMD filter including colour $BP-RP$ and its median error (calculated from \textit{Gaia} data and used as the filter width) as a function of apparent $G$.}
    \label{fig:kinematic_cmd_filters}
\end{figure}

We bin particles in $\phi_1$ and compute means and standard deviations of $\phi_2$, $\mu^*_{\alpha}$ and $\mu_{\delta}$ within each bin. The results are shown in Figure~\ref{fig:kinematic_cmd_filters} in blue points, based on which we create kinematic filters. The mean tracks are smoothed by fitting splines. The dispersions (or widths) are simply substituted by straight lines, from $1\degr$ and 0.5 mas~yr$^{-1}$ at $\phi_1=-60\degr$ to $5\degr$ and 1.0 mas~yr$^{-1}$ at $\phi_1=60\degr$, for $\phi_2$ and proper motions, respectively. These tracks and widths of filters are shown in red lines in Figure~\ref{fig:kinematic_cmd_filters}. To be conservative, the width lines are all above predictions from the model, meaning that the filtering windows will be a little wider. We use straight lines to avoid artifacts; for instance, $\mu_{\delta}$ dispersion of the model becomes narrower suddenly at $\phi_1=0\degr$ due to this part getting closer to the pole of Dec = $-90\degr$, which would cause less stars to be selected. We also fit splines to the $\phi_2$ model width $s_{\phi_2}$ (the green line in Figure~\ref{fig:kinematic_cmd_filters}) that will be used in the density modelling below. Here the spline fitting is achieved in this way: a pair of cubic splines with equidistant knots of 20$\degr$ represent the track and width as functions of $\phi_1$; the model particles around certain $\phi_1$ are assumed to be distributed following a Gaussian likelihood with parameters given by spline pairs; the track and width of the trial model can be fitted by sampling splines. 

\subsubsection{CMD Filter}

Although the spectroscopic metallicity of Indus is $-1.9$ dex \citep{2022ApJ...928...30L}, we choose to adopt an isochrone \citep[PARSEC;][]{2012MNRAS.427..127B} with age = 13 Gyr and [M/H] = $-1.5$ dex to serve as the filter. We find that the isochrone better matches with the location of the Indus stars in the CMD, compared to an isochrone with [M/H] = $-1.9$ dex that deviates from the Indus stars at the red giant branch. We note that the $-1.5$ dex isochrone only acts as the CMD filter here, and we still use $-1.9$ dex as the metallicity when creating luminosity function and synthetic stellar population of Indus below. The filter will be placed at corresponding distance moduli at different $\phi_1$, where the distance is based on the above trial model as well. To select stars around the isochrone, a dependence of colour $BP-RP$ error on apparent $G$ magnitude is derived by calculating colour error medians in $G$ bins (bin = 0.5 mag), which is referred to as ``colour width''. The final CMD filter is displayed in bottom of Figure~\ref{fig:kinematic_cmd_filters}.

\subsection{Filtered Maps} \label{subsec:filtering}

We apply the following filtering selections to \textit{Gaia} data:  
\begin{enumerate}
    \item $|$ $\phi_2$ data $-$ $\phi_2$ track $|$ $<$ $\phi_2$ width,
    \item $|$ $\mu^*_{\alpha}$ data $-$ $\mu^*_{\alpha}$ track $|$ $<$ $\mu^*_{\alpha}$ width,
    \item $|$ $\mu_{\delta}$ data $-$ $\mu_{\delta}$ track $|$ $<$ $\mu_{\delta}$ width,
    \item $|$ colour data $-$ isochrone colour $|$ $<$ 3 $\times$ colour width,
\end{enumerate}
where (i), (ii), and (iii) are performed against $\phi_1$, and (iv) against $G$. Note that filter width in (iv) is triple the colour width in order to obtain a relatively wider filtering window (similar to the customized larger dispersions in kinematic filters), and the isochrone will be adjusted at appropriate distance moduli for different $\phi_1$ bins (bin width = $3\degr$). For all filtering in the following, we further set up a parallax (in mas) cut of $\varpi - 3\sigma_{\varpi} <$ 1/distance (in kpc) where the distance is varying along $\phi_1$ as well. This cut removes foreground stars closer than the stream distance at $3\sigma$ confidence. However, it would have no substantial effects on the results, because most distant \textit{Gaia} sources do not possess sufficiently precise parallax measurements and will be retained.

\begin{figure*}
    \includegraphics[width=\linewidth]{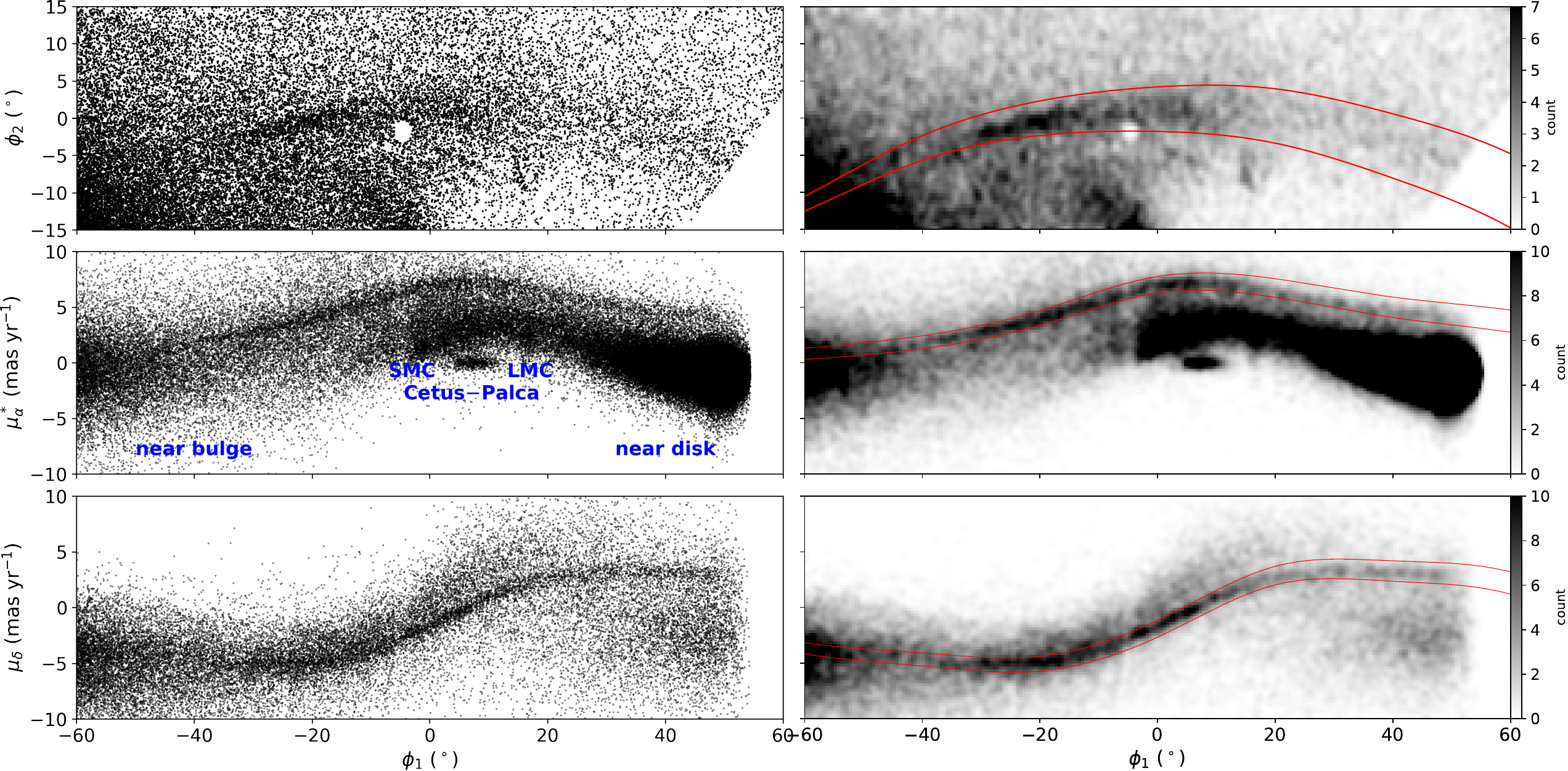}
    \includegraphics[width=\linewidth]{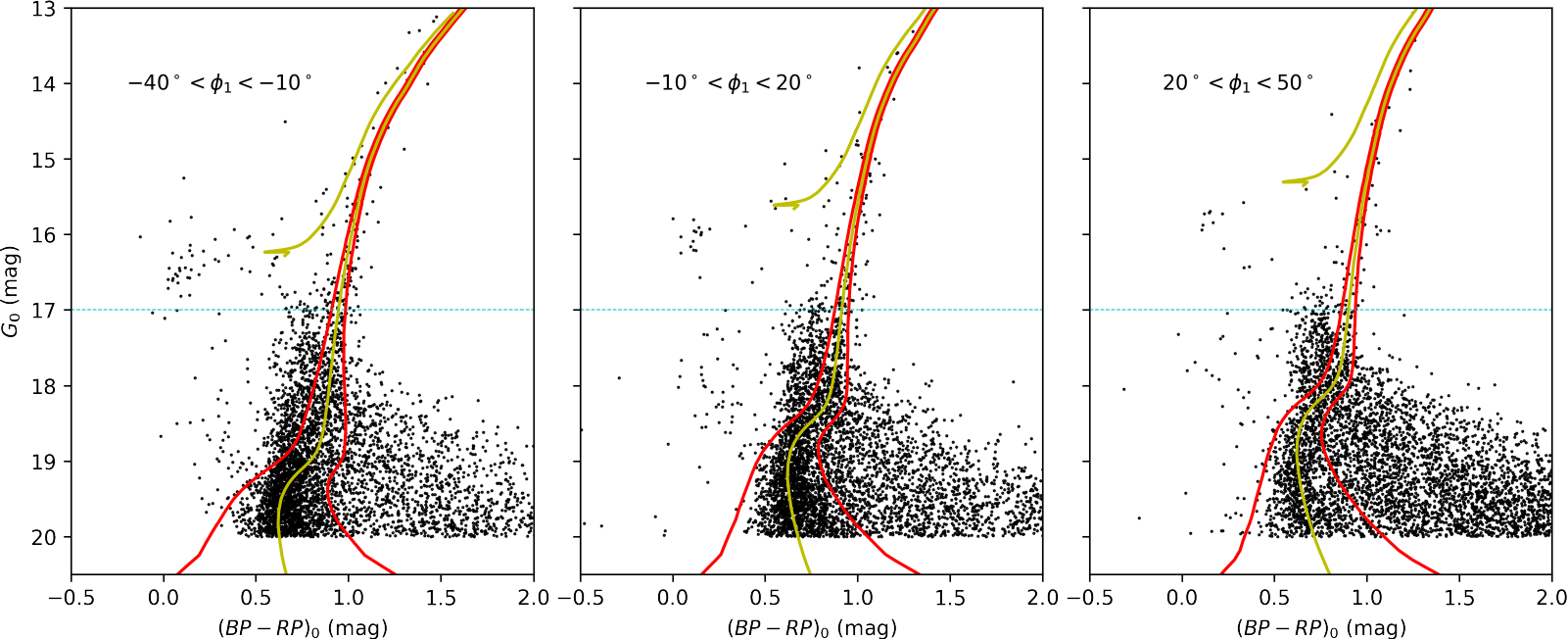}
    \caption{The first row presents the filtered map on the sky. Stars are shown in black dots in the left panel. The right panel is a binned two dimensional histogram of the left one. The $\phi_2$ filter is overplotted in the red curve as a comparison. The second and third rows are similar but for $\mu^*_{\alpha}$ and $\mu_{\delta}$. The last row presents the CMD of Indus in three segments. Yellow and red lines show the isochrone filters. Cyan dashed lines indicate $G_0=17$ mag, above which Indus dominates.}
    \label{fig:indus_kinematic_cmd}
\end{figure*}

The filtered maps are shown in Figure~\ref{fig:indus_kinematic_cmd}, with $\phi_2$, $\mu^*_{\alpha}$, $\mu_{\delta}$ and CMD spaces presented from top to bottom. Each dimension is the result of applying the other three filters. For example, the spatial map (top panel) is obtained by applying selections (ii), (iii), and (iv); while the CMD one is after applying selections (i), (ii), and (iii). 

The first row in Figure~\ref{fig:indus_kinematic_cmd} shows Indus in ($\phi_1$, $\phi_2$) sky coordinates. The stream motion is from negative to positive $\phi_1$, where the negative side points towards the Galactic bulge and the positive one goes close to the disk (data truncated at $b$ = $-10\degr$). Due to the distance gradient (see below), Indus is concentrated and prominent at the further side from $\phi_1$ $\sim$ $-40\degr$ to $10\degr$, while it is dispersed (wider) and indistinct at the nearer side of $\phi_1$ $>$ $10\degr$. The globular cluster NGC~362, at $\phi_1$ $\sim$ $-5\degr$, is masked due to its extra-tidal structure \citep{2019MNRAS.486.1667C}, and a small dense area at $\phi_1$ $\sim$ $15\degr$ corresponds to the LMC. The right column is a two dimensional histogram of the left one, where the red curve is the filter in Figure~\ref{fig:kinematic_cmd_filters}. 

The second row presents Indus in $\mu^*_{\alpha}$. A clear track is visible above field stars that is in agreement with the filter. Here, background stars mainly contain the bulge, SMC (Small Magellanic Cloud), a segment of Cetus-Palca debris \citep{2020ApJ...905..100C,2022ApJ...930..103Y}, LMC, and the disk. As a result, $\mu^*_{\alpha}$ can be used to filter out non-Indus stars efficiently. The third row gives similar plots but in $\mu_{\delta}$. Particularly at large negative $\phi_1$ (to the left of the plot), the stream is difficult to distinguish from bulge stars.

The last row slices Indus into three segments of $30\degr$ long in the CMD. Yellow lines show the isochrone (13 Gyr, $-$1.5 dex) at distance moduli of central $\phi_1$ for each segment, that is, $\phi_1$ = $-25\degr$, $5\degr$, and $35\degr$. Red lines around each isochrone mean three times the colour width in the CMD filter. Overall, the main sequence appears to be overdense, but is contaminated by field stars. The red giant and horizontal branches above $G_0=17$ mag are dominated by Indus, which match with the isochrone filter well.

\subsection{Gradients}

\begin{figure}
    \includegraphics[width=\columnwidth]{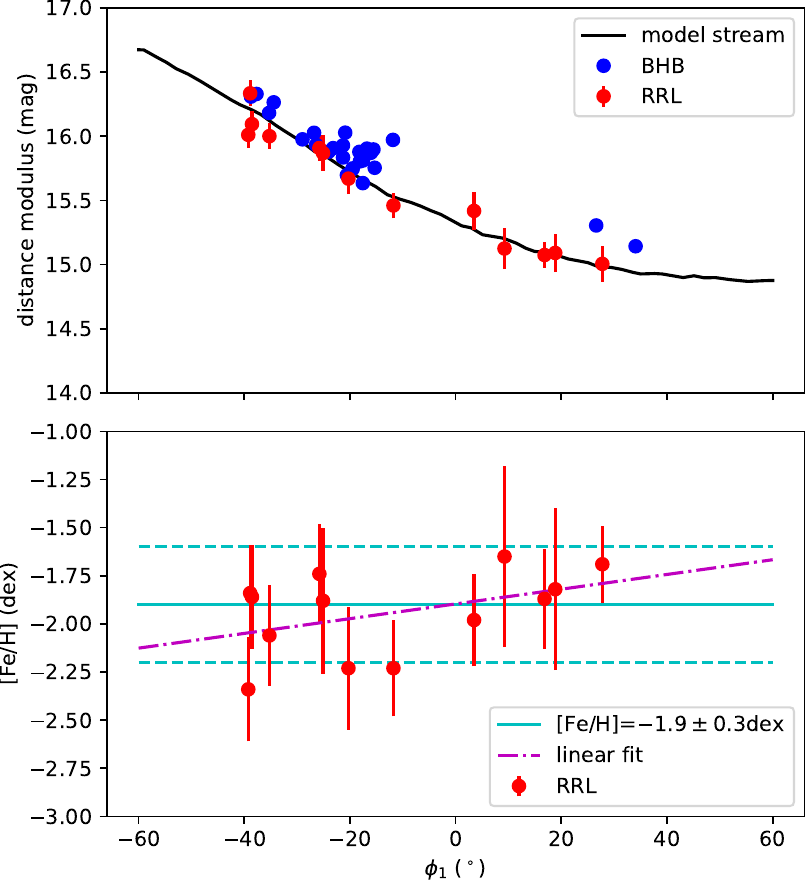}
    \caption{Distance modulus and metallicity gradients of the Indus stream. Red and blue points are crossmatched RRL and BHB stars in the CMD of Figure~\ref{fig:indus_kinematic_cmd}. The black line represents the distance of the trial model. Magenta and cyan lines show a linear fit and the mean and dispersion of metallicity of Indus reported in \citet{2022ApJ...928...30L}. }
    \label{fig:dm_feh}
\end{figure}

As there are a number of horizontal branch stars in the filtered CMD, we crossmatch them with the RR Lyrae (RRL) catalogue from \citet{2023ApJ...944...88L} to obtain their photometric metallicity and distance measurements, which are derived through calibrated period$-$Fourier~parameter$-$[Fe/H] relation and absolute~$G$~magnitude$-$[Fe/H] relation. We further supplement distances by crossmatching with the blue horizontal branch (BHB) star catalogue from \citet{2024A&A...690A.166A}. Their distance moduli are plotted against $\phi_1$ in Figure~\ref{fig:dm_feh}, along with the distance of the trial model stream. We observe that they are largely consistent, meaning it is feasible to employ the model as a guide. Additionally, we plot the photometric metallicity as a function of $\phi_1$. The mean metallicity of Indus $=-1.9$ dex along with its dispersion = 0.3 dex \citep{2022ApJ...928...30L} is indicated as cyan solid and dashed lines. A linear fit to stars is shown by the magenta line, from which a clear gradient is observed. A slightly higher metallicity towards positive $\phi_1$ probably implies that the more negative $\phi_1$ section of the stream was once the outskirts of Indus galaxy, with the dwarf core located toward the more positive end.

\subsection{Stellar Mass}

It is possible to estimate the stellar mass of Indus based on the number of likely members we detect. We use filtered stars brighter than $G_0$ = 17 mag in the CMD of Figure~\ref{fig:indus_kinematic_cmd}, which are mainly made of Indus plus a few background stars (denoted as $n_{s+b}$). To assess the background, we apply filtering selections (i), (ii), and (iii) of Section~\ref{subsec:filtering} again, but move the ``$\phi_2$ track'' by $10\degr$ along the positive $\phi_2$ axis. The filtered stars of this parallel region above $G_0$ = 17 mag in CMD can be an estimate of the background (denoted as $n_b$). Thus $n_{s+b}-n_b$ is the number of stream stars here. However, this value should roughly correspond to 68\% of the total number since we use 1$\sigma$ width in the $\phi_2$ filter in Figure~\ref{fig:kinematic_cmd_filters} ($\mu^*_\alpha$ and $\mu_\delta$ dispersions are designed to be much wider, and we expect little incompleteness). As a result, we estimate the actual number to be ($n_{s+b}-n_b$)/0.68 $\simeq$ 419 stars. After correcting apparent $G$ magnitudes into absolute magnitudes $M_G$ using the distance modulus from Figure~\ref{fig:dm_feh}, these stars are located between $-3<M_G<2$ mag. Comparing this to a luminosity function corresponding to 13 Gyr and $-1.9$ dex, a mass of $2 \times 10^5$ M$_{\sun}$ is required to match the observed number of stars in this $M_G$ range, giving 422 stars. However, this is a lower limit on the stream mass because there might be an undetected portion of Indus fading into the Galactic bulge or disk, hard to distinguish for now. The fact of Indus being a dwarf galaxy in the past also allows us to estimate its stellar mass using the mass$-$metallicity relation \citep[Equation 4 in][]{2013ApJ...779..102K} for dwarf galaxies of the local group. The metallicity of $-1.9$ dex also gives a stellar mass of $2 \times 10^5$ M$_{\sun}$, consistent with the value from the luminosity function. 

\section{Morphology characterization} \label{sec:density}

Through the matched-filter map in Figure~\ref{fig:indus_kinematic_cmd}, we can observe the underlying density fluctuations in Indus by eye. In order to quantify the stream's morphology, we perform a density fit to the data.

\subsection{Density Fitting} \label{subsec:pymc}

We employ the density-fitting technique developed by \citet{2017MNRAS.470...60E} and \citet{2019MNRAS.485.4726K}; see also \citet{2021ApJ...911..149L,2022MNRAS.514.1757P,2022AJ....163...18F,2023MNRAS.521.4936K,2025arXiv250621410C}. The spatial map ($\phi_1$, $\phi_2$) of Figure~\ref{fig:indus_kinematic_cmd} is decomposed into two components: a stream and a background. The stream component is assumed to have a Gaussian cross section at any $\phi_1$:

\begin{equation}
\rho_s(\phi_1,\phi_2) = \frac{\mathrm{exp}(I(\phi_1))}{\mathrm{exp}(S(\phi_1))\sqrt{2\pi}}
\mathrm{exp} \left( -\frac{1}{2}\left( \frac{\phi_2-\Phi_2(\phi_1)}{\mathrm{exp}(S(\phi_1))} \right)^2 \right) 
\label{eq:rho_s}
\end{equation}
where $I(\phi_1)$ and $S(\phi_1)$ are logarithms of stream intensity and width, while $\Phi_2(\phi_1)$ is stream trajectory. The background is described as a simple polynomial:
\begin{equation}
\rho_b(\phi_1,\phi_2) = \mathrm{exp}(a(\phi_1)\phi^2_2+b(\phi_1)\phi_2+c(\phi_1)) .
\label{eq:rho_b}
\end{equation}
The total density is given by $\rho_s+\rho_b$.

The model is fitted to a two-dimensional histogram of the filtered data on the sky between $-40\degr < \phi_1 < 50\degr$ and $-15\degr < \phi_2 < 15\degr$ (the pixel size is $0.5\degr\times0.5\degr$), with LMC masked in $13\degr<\phi_1<20\degr$ and $-10\degr<\phi_2<-3\degr$. We use cubic splines to represent functions $I(\phi_1)$, $\Phi_2(\phi_1)$, $S(\phi_1)$, $a(\phi_1)$, $b(\phi_1)$, and $c(\phi_1)$. We manually place the spline knots for the stream component equidistantly at $\phi_1=$ $(-40\degr, -30\degr, -20\degr, -10\degr, 0\degr, 10\degr, 20\degr, 30\degr, 40\degr, 50\degr)$, and use the Bayesian optimization \texttt{gp\_minimize} implemented in \texttt{scikit-optimize}\footnote{\url{https://doi.org/10.5281/zenodo.1207017}.} to search for the best number of splines for the background functions by evaluating the Akaike information criterion (AIC),
\begin{equation}
\mathrm{AIC} = 2k - 2 \mathrm{ln} \mathcal{L}  .
\end{equation}
Here $k$ is the number of all spline coefficients and $\mathcal{L}$ is the likelihood,
\begin{equation}
\begin{split}
\mathcal{L} = & \mathrm{Poisson}(n_\mathrm{data}|n_\mathrm{model}) \\
            & \times \mathrm{Normal(\Phi_2|\phi_2 \  track,1\degr)}  \\
            & \times \mathrm{Normal}(S|\mathrm{ln}s_{\phi_2},0.5\degr) .
\end{split}
\label{eq:L}
\end{equation}
A random sample of spline coefficients will translate into $I(\phi_1)$, $\Phi_2(\phi_1)$, $S(\phi_1)$, $a(\phi_1)$, $b(\phi_1)$, and $c(\phi_1)$, which further produce a model density map through Equation~\ref{eq:rho_s} and \ref{eq:rho_b}. The likelihood first contains a Poisson function taking in the number counts in the data ($n_\mathrm{data}$) given the model counts ($n_\mathrm{model}$), which is enabled by the ``unweighted'' matched filter above. Thus all exponentials in Equations~\ref{eq:rho_s} and \ref{eq:rho_b} are aimed to make sure the corresponding quantities are $>0$. Two constraints of priors are placed on stream trajectory and width. We let $\Phi_2(\phi_1)$ and $S(\phi_1)$ follow Gaussian distributions with parameters $(\phi_2 \  \mathrm{track}, 1\degr)$ and $(\mathrm{ln}s_{\phi_2}, 0.5\degr)$, respectively, where $\phi_2$ track and $s_{\phi_2}$ are shown with red and green lines in first two panels of Figure~\ref{fig:kinematic_cmd_filters}. These two priors are introduced to make sure that the stream component does not deviate too far from the model predictions.

\texttt{gp\_minimize} will explore the numbers of spline bases for $a(\phi_1)$, $b(\phi_1)$, and $c(\phi_1)$ all from 5 to 30. For a given number set, we first obtain the best-fit coefficients by minimizing $-\mathrm{ln} \mathcal{L}$ using \texttt{scipy.optimize.minimize} \citep{2020SciPy-NMeth}, and then return the AIC value to \texttt{gp\_minimize}. The first term $2k$ in AIC acts as penalty to avoid overfitting because $-2\mathrm{ln}\mathcal{L}$ always favours more numbers of splines (the model will be more and more like the data). As mentioned previously, spline knots for the stream functions $I(\phi_1)$, $\Phi_2(\phi_1)$ and $S(\phi_1)$ are manually defined. Using just AIC to set their spline numbers would reduce them all to 5, the minimum allowable number of the searching range. It is because 5 splines are enough to describe an overall trends of three functions and penalty comes more than the likelihood refinement in AIC if the number further increases. However, this is not enough to capture smaller-scale structures in the stream. The optimal numbers found by \texttt{gp\_minimize} are 18, 11, and 16 for $a(\phi_1)$, $b(\phi_1)$, and $c(\phi_1)$, respectively.

In total, the parameters to be determined are 81 spline coefficients. To obtain posterior distributions of results, we apply the Hamiltonian Monte Carlo No-U-Turn Sampler \citep[NUTS;][]{Hoffman2014} built in \texttt{pymc} \citep{AbrilPla2023} to efficiently sample the high-dimensional parameter space. The likelihood is still $\mathcal{L}$ in Equation~\ref{eq:L}, and the prior for all coefficients is a broad Gaussian, $\mathrm{Normal}(0, 10)$. The sampling is run in four parallel chains that contain 1000 burn-in and 1000 samples each. The Gelman–Rubin \^{R} = 1.0 indicates that four sampling chains reach the convergence.

\subsection{Quantitative Density}

\begin{figure}
    \includegraphics[width=\columnwidth]{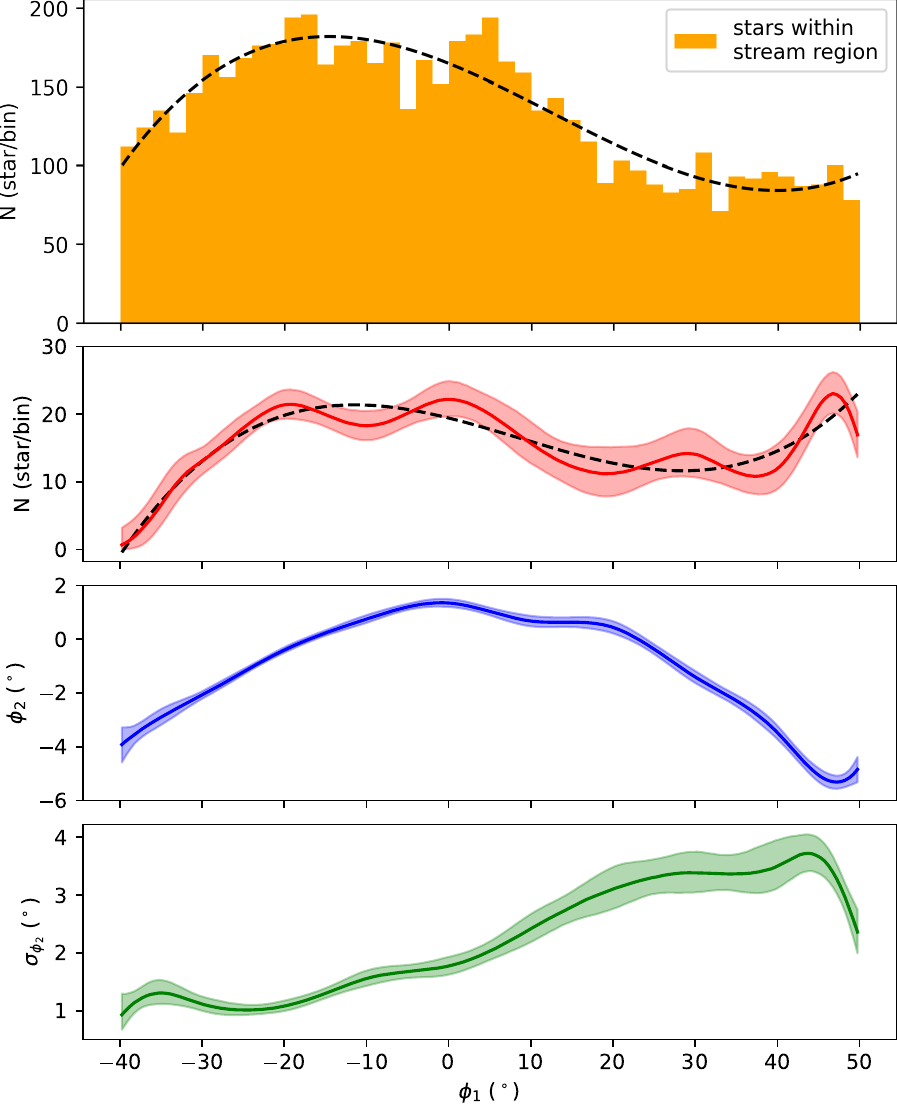}
    \caption{The top panel is a histogram of all stars (stream $+$ background) located in the stream region between the red lines in the upper-right panel of Figure~\ref{fig:indus_kinematic_cmd}. The rest three panels are measurements of stream density N (plotted as $\mathrm{exp}(I(\phi_1))/0.5$), stream trajectory $\phi_2$ (i.e. $\Phi_2(\phi_1)$) and stream width $\sigma_{\phi_2}$ ($\mathrm{exp(S(\phi_1))}$) in Equation~\ref{eq:rho_s}. Dividing $\mathrm{exp}(I(\phi_1))$ by $\phi_2$ bin width $=0.5\degr$ is used to derive total star counts within each $\phi_1$ bin. The  Solid lines and bands show results at 16th, 50th, and 84th percentiles. Additionally, black dashed lines are third-order polynomial fits to respective profiles, used to normalize densities to compute power spectra later.}
    \label{fig:density_model}
\end{figure}

Figure~\ref{fig:density_model} presents measurements of stream density ($\mathrm{exp}(I(\phi_1))/0.5$), stream trajectory ($\Phi_2(\phi_1)$) and stream width ($\mathrm{exp(S(\phi_1))}$), where dividing by $\phi_2$ bin width $=0.5\degr$ is used to derive total star counts within each $\phi_1$ bin. The solid lines and bands show results at 16th, 50th, and 84th percentiles. On the top is a histogram of all stars (stream $+$ background) located in the stream region, that is, between the red lines in the upper-right panel of Figure~\ref{fig:indus_kinematic_cmd}. This histogram is aimed to be a direct comparison with the stream density profile $I(\phi_1)$ we have measured. The background is supposed to vary gradually, and we expect the histogram to reflect variations of the stream itself. In general, we can note similar density trends from the first two panels, both of which reveal some episodic fluctuations in stellar counts. Specifically, broad gaps are present at about $\phi_1$ = $-10\degr$, $20\degr$ and $40\degr$, while on both sides of gaps are several peaks roughly at $\phi_1$ = $-20\degr$, $0\degr$, $30\degr$ and $50\degr$. We consider that these peaks are due to Indus rather than any other substructures, which can be justified with Figure~\ref{fig:indus_kinematic_cmd}. After selections (i), (iii), and (iv), the stream is well distinguished from background stars in $\mu^*_{\alpha}$. This reflects that for the stream area on the sky after (iii) and (iv), it still consists of Indus and some substructures (like SMC, Cetus-Palca, LMC). The density profiles here in Figure~\ref{fig:density_model} are obtained after applying (ii), (iii), and (iv). It means that we further clean up contaminants using (ii), and now the stream sky area only contains Indus and a few background but no major substructures. That area is principally where we derive the longitudinal density. The stream fades gradually to the negative end of $-40\degr$ but seems to continue beyond the positive end which comes close to the Galactic disk at $b=-10\degr$. The third and fourth panels display the stream trajectory and width. Indus becomes broader angularly when sliding from negative to positive $\phi_1$ due to it approaching closer to us. This measured increase in the width is closely consistent with $s_{\phi_2}$ from the trial model, which is reasonable since $s_{\phi_2}$ is included as the prior during the fitting. We note that it is necessary to include a prior to fit $S(\phi_1)$ because $S(\phi_1)$ in exponential acts as $\sigma$ in Gaussian, and a random big or small $S(\phi_1)$ will result in an extreme $\sigma$, making the fitting not converge.

\begin{figure}
    \includegraphics[width=\columnwidth]{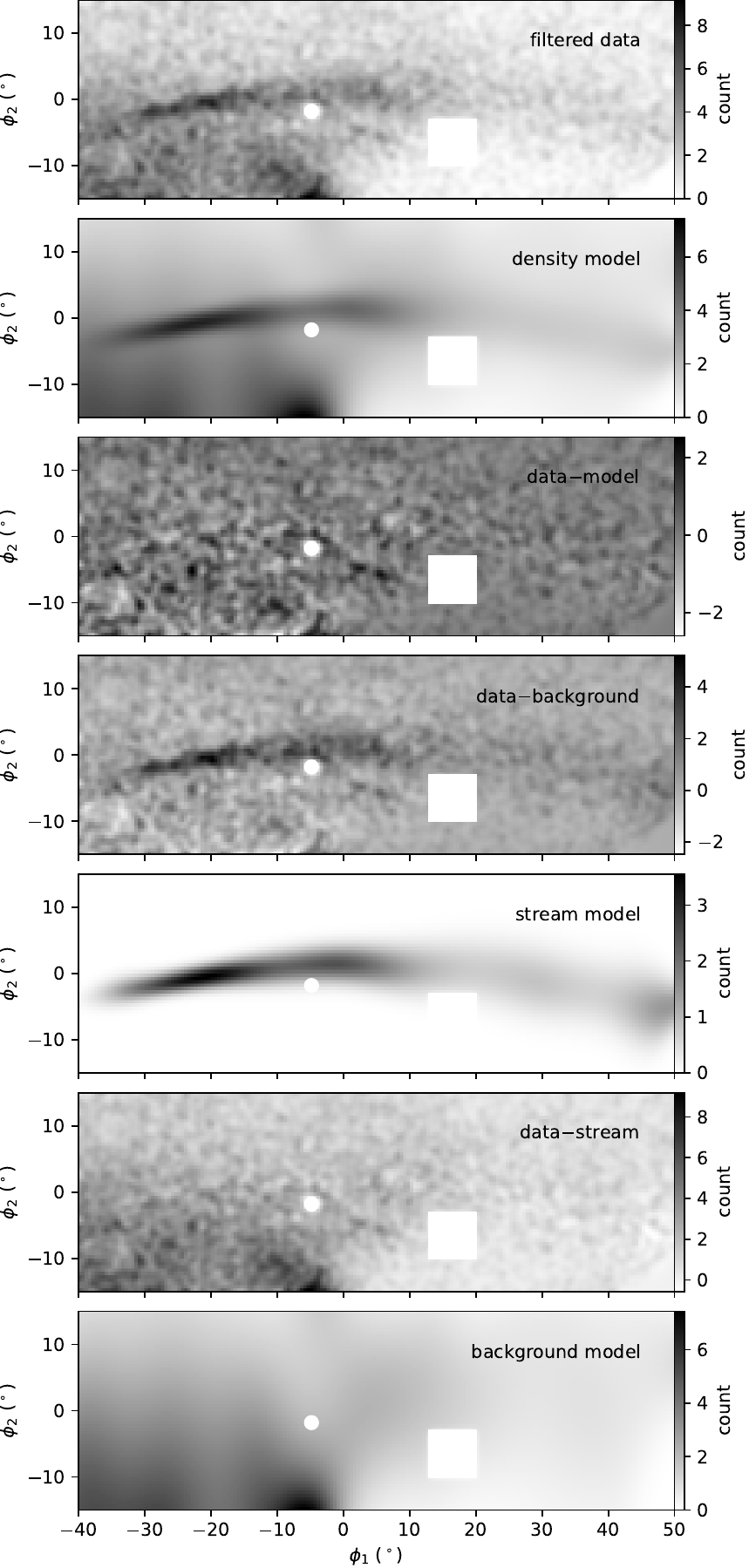}
    \caption{The top panel is the filtered spatial map from Figure~\ref{fig:indus_kinematic_cmd}. The second panel shows the density model constructed from the best-fit $I(\phi_1)$, $\Phi_2(\phi_1)$, $S(\phi_1)$, $a(\phi_1)$, $b(\phi_1)$, and $c(\phi_1)$ in Equation~\ref{eq:rho_s} and Equation~\ref{eq:rho_b}. The third panel is the residual between the first two panels. The fourth panel presents data map after removing the fitted background. The fifth one shows the pure stream component in the best-fit density. Similarly, the last two panels show the data after subtracting the stream model and the pure background model. The NGC~362 and LMC are masked with blank circle and square.}
    \label{fig:density_comparison}
\end{figure}

Figure~\ref{fig:density_comparison} shows several comparison of density maps between the model and the data. From the top to bottom are the original filtered map, the total fitted density, the residual between them, the data after removing the fitted background, the pure fitted stream, the data after removing the fitted stream, and the pure fitted background. The globular cluster NGC~362 and LMC are masked here with the blank circle and square. By looking at the first two panels, it is noted that the density model successfully reproduces the major morphologies existing in the data map. The negligible residual (third) indicates that the model is well matched to the data. The density fluctuations are presented more clearly in the stream-only panels (fourth and fifth) in the absence of the background component. The stream signals nearly disappear from the data after subtracting the stream model (sixth), showing agreements with the background model on the bottom. These visual comparisons represent the success of the density map fitting.

\subsection{Another Gap}

\begin{figure}
    \includegraphics[width=\columnwidth]{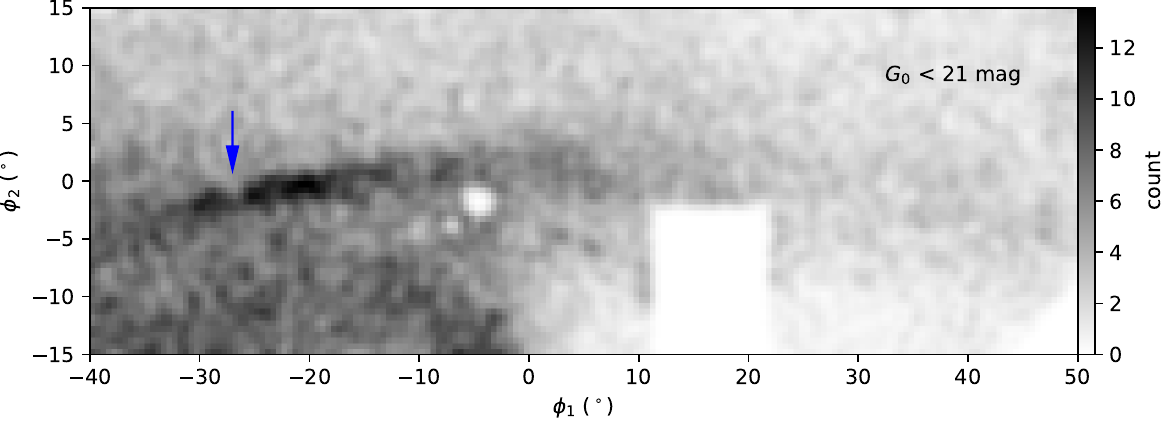}
    \caption{The spatial filtered map similar to Figure~\ref{fig:indus_kinematic_cmd} but using all stars of $G_0$ $<$ 21 mag. The blue arrow marks a tentative gap of the stream.}
    \label{fig:gap_G21mag}
\end{figure}

All above ``main'' results are based on the \textit{Gaia} data that are limited to stars brighter than $G_0$ = 20 mag, as mentioned in Section~\ref{sec:data}. However, due to the distance gradient of Indus (Figure~\ref{fig:dm_feh}), the flat apparent magnitude cut $G_0 > 20$ mag will preferentially discard stream stars on the negative $\phi_1$ side (more distant from us). If we remake the spatial map using all stars brighter than $G_0 = 21$ mag by applying the same filtering selections (ii), (iii), and (iv) in Section~\ref{subsec:filtering}, we expect to increase the contrast on the left side (but probably blur the right side with more contamination); this is shown in Figure~\ref{fig:gap_G21mag}. More of LMC stars are included now and it is masked with a bigger rectangle. It is seen that the overall stream morphology does not change too much, but there arises a sharp discontinuity to the leftmost side around $\phi_1$ = $-27\degr$ marked with a blue arrow. We note that there is already a subtle sign of this gap if we zoom in Figure~\ref{fig:indus_kinematic_cmd}, but it is rather obscure. Now including as many stream stars as possible makes it more prominent in Figure~\ref{fig:gap_G21mag}. This discontinuous feature also seemingly appears in \citet{2018ApJ...862..114S} (Dec $\sim$ $-60\degr$ in their Figure~5). We tend to believe that the gap at $\phi_1=-27\degr$ is real because what we have done is just to further select fainter stars in filtering windows, for which astrometric and photometric measurements might be more uncertain. Out of them, field stars are supposed to populate everywhere spatially, while the gap feature within the stream region should mainly be made of Indus stars. We expect that it can be verified by future \textit{Gaia} DR4 or DR5 in which fainter stars' errors would be revised.

\subsection{Excluding Artifacts}

\begin{figure}
    \includegraphics[width=\columnwidth]{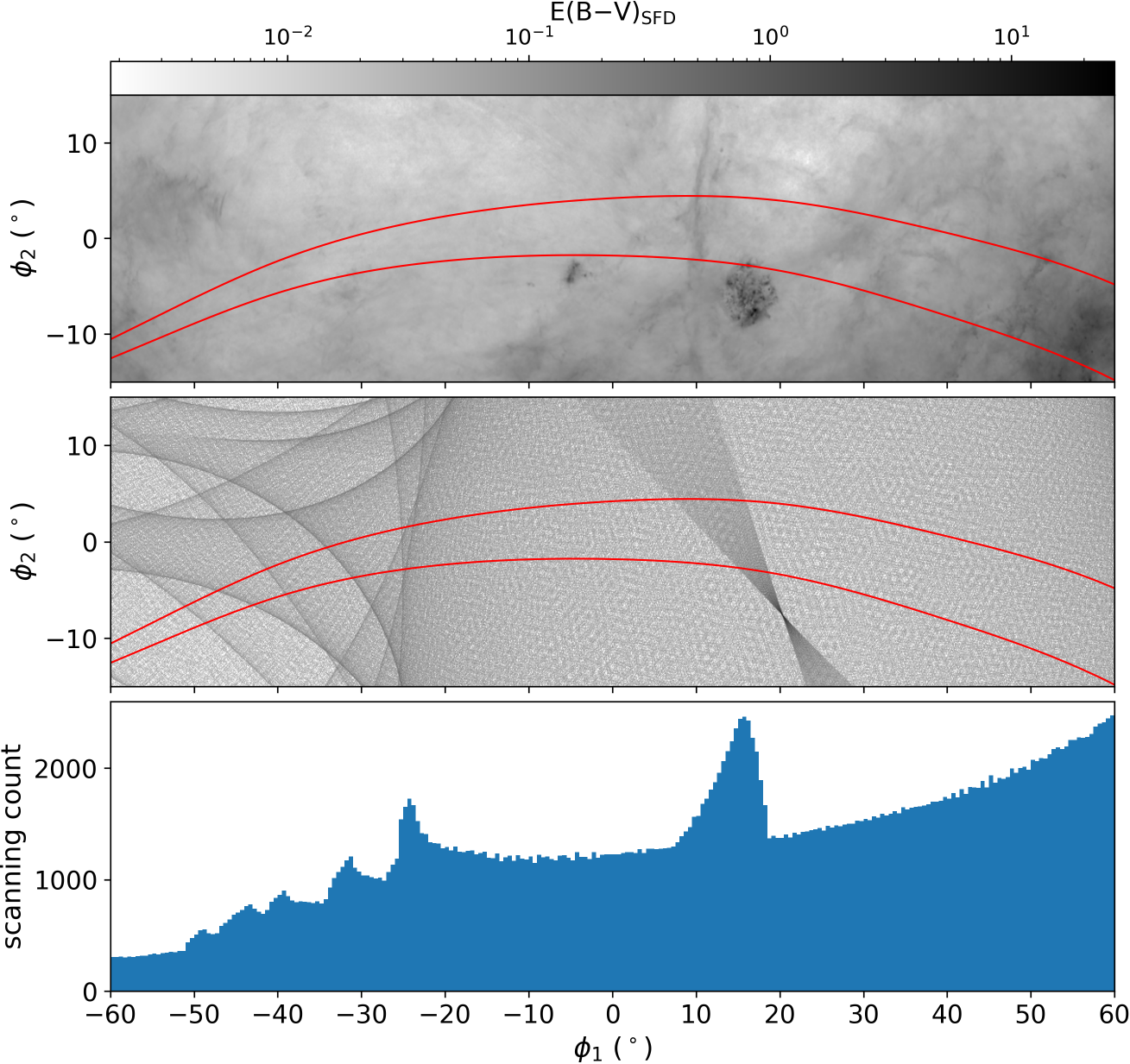}
    \caption{The top panel shows interstellar dust map from \citet{1998ApJ...500..525S} where two darker regions correspond to SMC and LMC. The middle panel shows \textit{Gaia} scanning pattern where darker colours mean more observation times and vice versa. The red lines indicate the spatial matched filters. The bottom panel is a histogram against $\phi_1$ of scanning times between the red lines in the middle panel.}
    \label{fig:dust_scan}
\end{figure}

To confirm that the density variations of Figure~\ref{fig:density_model} are not artifacts, we compare the stream detection with the dust map \citep{1998ApJ...500..525S} (top panel) and the scanning pattern of the \textit{Gaia} satellite (middle and bottom panels) in Figure~\ref{fig:dust_scan}. The spatial matched filter is overplotted in red lines as a reference. The first concern is whether the stream spans across any major interstellar structures. If a stream encounters a heavily dusty region, it might lead to inaccurate extinction corrections and cause some offsets in the CMD for target stars. Then the matched filter might overlook them partially, which produces an artificial underdensity. We dispel this concern by looking at the dust distribution. On the one hand, we can trust extinction corrections for most of the area (except SMC and LMC), including the majority of Indus because it is clear from the top panel that there is no high-spatial-frequency structure in the dust map along the Indus stream track. Thus these stream stars are properly captured in the CMD filtering. On the other hand, although dereddening can be problematic in the direction of SMC and LMC, we consider that the density measurement is not heavily affected. SMC covers a tiny sky extent, wherein star counts of these pixels play a very limited role in the total likelihood calculation. Star counts in the LMC region are not taken into account (masked) such that a truncated Gaussian is fitted in practice here. Hence the interstellar dust should not bias the Indus detection. The rest two panels present the \textit{Gaia} scanning pattern, where the bottom histogram indicates scanning amounts falling between the red lines. It is seen that the density variations we measure in Figure~\ref{fig:density_model} do not reflect inhomogeneities of \textit{Gaia} observations. The peaks in Indus do not coincide with darker shadows (more observation times) around $\phi_1 \sim 15\degr$ or $<$ $-25\degr$, and the gaps are located in homogenous areas. Therefore, we conclude that the Indus density fluctuation should be real.

\section{Dynamic modelling} \label{sec:model}

The density fitting reveals episodic gaps and peaks in Indus, which motivates us to investigate their origins. As introduced in Section~\ref{sec:intro}, it could be in relation to internal dynamics like epicyclic motions, or external perturbations by baryonic structures and dark matter subhalos. Our preliminary design is to first inspect the internal one, that is, whether a natural tidal disruption can produce similar density fluctuations; if the answer is yes, we will evaluate the necessity of exploring external possibilities. On the other hand, if the answer is no, the simulation should leave us a moderately smooth model stream with a relatively flat density, which will allow us to further incorporate external effects.

\subsection{N-body Settings}

Prior to modelling the disruption of Indus, we refine the stream's orbit fitting following the method in Section~\ref{subsec:k-filter}. The procedure is mostly the same except for two changes: (1) apart from Indus members from $S^5$, we also incorporate the fitted stream track and width of Figure~\ref{fig:density_model} in the Gaussian likelihood, and (2) we anchor the progenitor at $\phi_1=50\degr$ because a slight positive metallicity gradient in Figure~\ref{fig:dm_feh} is observed implying that Indus' inner core might be on the higher-metallicity side. The best-fit results for the progenitor position are ($\phi_1$, $\phi_2$, $d_{\sun}$, $\mu^*_{\alpha}$, $\mu_{\delta}$, $V_r$) $=$ ($50\degr$, $-4.52^{+0.23}_{-0.22}\degr$ $9.50 \pm 0.09$ kpc, $4.53^{+0.05}_{-0.04}$ mas~yr$^{-1}$, $3.01 \pm 0.03$ mas~yr$^{-1}$, $227.42^{+1.19}_{-1.26}$ km~s$^{-1}$).

Since the Indus progenitor was a dwarf galaxy, we need to contain both stellar and dark matter components. We perform N-body simulations using \texttt{GADGET-4} \citep{2021MNRAS.506.2871S} to account for their internal gravity. Stars are described with a Plummer sphere \citep{1911MNRAS..71..460P} where its mass is adopted to be $2 \times 10^5$ M$_{\sun}$ (the scale radius is mentioned below). We use $10^5$ stellar particles and a softening of 5 pc. The dark halo is represented by a generalized NFW form \citep{1997ApJ...490..493N}, 
\begin{equation}
\rho(r) = \rho_0 \left( \frac{r}{r_s} \right)^{-\gamma}
\left( 1+ \frac{r}{r_s} \right)^{\gamma - 3} 
\mathrm{exp}\left[ -\left( \frac{r}{r_\mathrm{cut}} \right)^2 \right] ,
\label{eq:nfw}
\end{equation}
where $\rho_0$ is the density normalization, $r_s$ is the scale radius, $\gamma$ is the inner slope, and $r_\mathrm{cut}$ is the cutoff radius. A stellar mass of $2 \times 10^5$ M$_{\sun}$ roughly corresponds to a halo mass of $\sim 3 \times 10^8$ M$_{\sun}$ according to \citet{2017MNRAS.467.2019R}. In practice, we do not realize the dark halo through $\rho_0$, but directly set the total halo mass as that value, and $\rho_0$ can be determined internally. We choose $r_s$ = 3 kpc and $r_\mathrm{cut}$ = 7 kpc. These two radii serve to fine-tune the system's velocity dispersion such that the present-day streams have dispersions similar to that derived from observations (see below). Although this $r_s$ is bigger than usual extant dwarfs, we expect it to mimic a puffed-up status for Indus before disruption. $r_\mathrm{cut}$ is simply a radius roughly half way of the pericentre distance, to prevent particles from going beyond the other side of the Galactic centre. As for the inner slope of the profile, $\gamma$ = 1 corresponds to a ``cuspy'' halo with the density increasing steeply at small radii, while $\gamma$ = 0 has a ``cored'' halo namely a flat central profile. Including both enables us to see how central halo profiles affect the final density fluctuations for Indus. We use $10^5$ dark matter particles with a softening of 0.1 kpc. In addition, we add another contrasting test in which there are only stars following a Plummer profile, in order to see what the stream would look like if there is no dark matter halo. Therefore, our modelling consists of three cases: CuspyHalo, $10^5$ stars in a Plummer sphere with a scale radius of 0.3 kpc and $10^5$ dark matter particles in a cuspy NFW halo; CoredHalo, $10^5$ stars in a Plummer sphere with a scale radius of 0.5 kpc and $10^5$ dark matter particles in a cored NFW halo; StarsOnly, $10^5$ stars in a Plummer sphere with a scale radius of 0.75 kpc. These radii (for both stars and dark halo) are somewhat arbitrary but are aimed to yield a velocity dispersion for the final model stream close to 7 km s$^{-1}$ that is measured from data \citep{2022ApJ...928...30L}. 

To generate initial conditions, we make use of \texttt{AGAMA} by randomly drawing particles from spherical distribution functions constructed from the density profiles above. It is very uncertain when the progenitor system moved in close enough within the gravitational field to undergo disruption \citep[e.g.,][]{2000MNRAS.314..468J}. On the one hand, Indus is located in the inner halo ($<20$ kpc) implying that the dwarf has been wandering in the Milky Way for a sufficient time. Hence we want to model it as long as possible. On the other hand, the current Galactic potential could be different if we trace back too far in time. Here, we adopt 5 Gyr to be the integration time, which is also the maximum interval supported by the \texttt{potMWLMC} we use as in Section~\ref{subsec:k-filter}. Under the \texttt{potMWLMC} potential, we first rewind the best-fit orbit back to 5 Gyr ago, and locate an apocentre near that time, which happens at $-4.847$ Gyr. We use the apocentre as the beginning point of the stream disruption by displacing initial conditions back there. The internal mutual gravity between all particles is calculated by \texttt{GADGET-4}, and the external force of the Milky Way as well as LMC is still from the \texttt{potMWLMC}. We note that the \texttt{AGAMA} potential can be used as a source of external gravity with some adjustments to  \texttt{GADGET-4}\footnote{See \url{https://github.com/GalacticDynamics-Oxford/Agama/blob/master/py/example_nbody_simulation.py}.}.

\subsection{Longitudinal Densities}

\begin{figure}
    \includegraphics[width=\columnwidth]{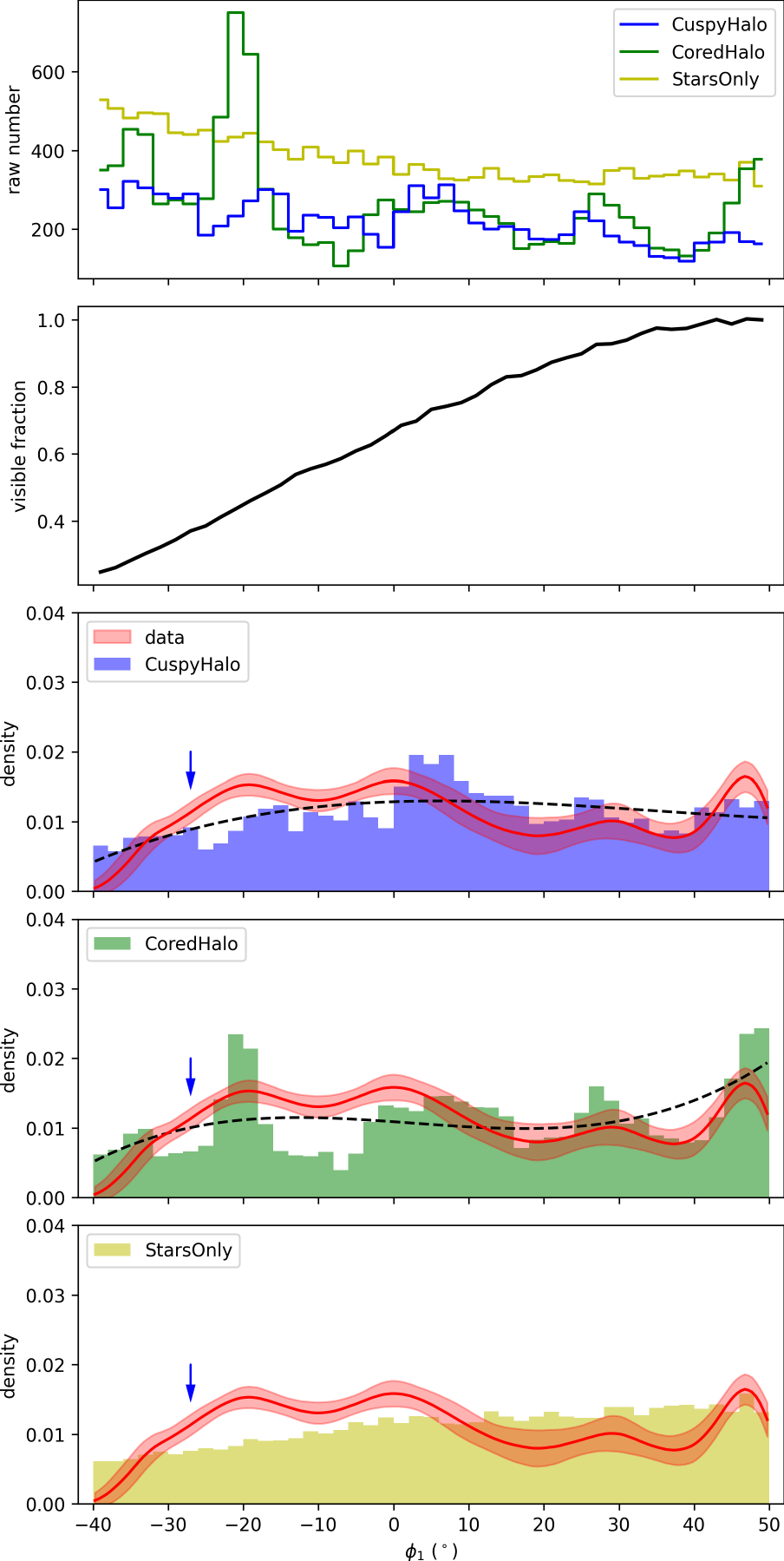}
    \caption{The first panel shows binned ($\phi_1$ bin = $2\degr$) raw numbers of stellar particles in different simulations. Blue, green, and yellow steps correspond to the N-body modelling for Indus with a cuspy halo (CuspyHalo), a cored halo (CoredHalo), and only stars (StarsOnly), respectively. The second panel shows the relative visible fraction of the Indus synthetic population due to its distance gradient and the magnitude cut $G_0<20$ mag. The blue, green, and yellow histograms in the remaining three panels show processed density profiles of model streams generated by the three types of N-body simulations of Indus. Black dashed lines are third-order polynomial fits to model densities of CuspyHalo and CoredHalo, used to divide density profiles to compute power spectra later. The red band is measured from data as in Figure~\ref{fig:density_model}. The blue arrow points out the location of an underlying gap that becomes more clearly visible when using all stars with $G_0<21$ mag as in Figure~\ref{fig:gap_G21mag}.}
    \label{fig:density_phi1}
\end{figure}

We show the present-day density profiles of Indus from N-body simulations in Figure~\ref{fig:density_phi1}. After nearly 5 Gyr of disruption, the stream is far longer than detected, containing $\sim$ 5 wraps or more. We only focus on the central wrap surrounding the progenitor spanning between $\phi_1$ = $-40\degr$ and $50\degr$. We also note that dark matter particles are quite dispersed across the sky and Figure~\ref{fig:density_phi1} is just displaying stars.

The first panel displays histograms of raw numbers of stellar particles in different simulations, where $\phi_1$ bins have a width of $2\degr$. Tidal disruptions finally leave fewer stars and stronger density variations for CuspyHalo and CoredHalo. By contrast, the StarsOnly disruption produces a fairly flat density and retains more stars here. Given that the density measurement in Figure~\ref{fig:density_model} is the result after proper-motion filtering and $G_0<20$ mag limit, we process simulation results in a similar way. We first apply proper-motion selections (ii) and (iii). To deal with the incompleteness of fainter stars due to the distance gradient of the stream, we generate a synthetic stellar population of Indus using age, [M/H], and mass = 13 Gyr, $-1.9$ dex, and $2 \times 10^5$ M$_{\sun}$, respectively. For each $\phi_1$ bin (width $=2\degr$) from $-40\degr$ to $50\degr$, the population is adjusted to the appropriate distance modulus in the CMD, and synthetic counts of stars brighter than $G_0=20$ mag are recorded. We can denote the visible count of $i$th bin as $m_i$, and the second panel of Figure~\ref{fig:density_phi1} shows $m_i/m_{-1}$ where $m_{-1}$ is the maximum of the last bin at $\phi_1$ = $50\degr$. If the total population is $n$, and $x_i$ represents the raw actual count of Indus stars in $i$th bin, we expect to observe $x_i m_i/n$ stars in each bin, which should correspond to what we measure in the second panel of Figure~\ref{fig:density_model}. To make it easy to compare different profiles, we normalize histograms of stellar counts through 
\begin{equation}
\frac{x_i m_i/n}{\mathrm{bin} \  \sum(x_i m_i/n)} 
= \frac{x_i m_i}{\mathrm{bin} \  \sum(x_i m_i)}
\label{eq:normalize}
\end{equation}
where bin = $2\degr$. For the profile measured from the data, the former format is applied as $x_i m_i/n$ is directly the number we observe. For profiles of the models, we use the latter format where $x_i$ is binned stellar counts from N-body simulations after proper-motion selections, and $m_i$ is already prepared above. Thus $n$ is not needed to be known. The resulting density profiles are shown in the lower three panels in Figure~\ref{fig:density_phi1}.

The third panel presents the profile of the model generated by assuming a cuspy dark matter halo for the Indus dwarf. We note that multiple peaks arise at $\phi_1 \sim$ $-15\degr$, $5\degr$, $25\degr$, and $45\degr$, which are possibly matched to these peaks detected in data at $-20\degr$, $0\degr$, $30\degr$, and $45\degr$ accordingly. There are also gaps in between around $-25\degr$, $-10\degr$, $20\degr$, and $40\degr$, similar to those in data if we further take into account the underlying gap at $-27\degr$ marked with the blue arrow that shows up in Figure~\ref{fig:gap_G21mag}. Therefore, even though there are minor location offsets, the numbers of peaks and gaps match well between the model and data. Similar conclusions can be drawn for the model in the presence of a cored halo as displayed in the fourth panel. The low-density regions in this model are located near the gaps in the data, namely, at $-27\degr$, $-10\degr$, $20\degr$, and $40\degr$. On both sides are density peaks located comparable to the measurement from data. It is worth noting that some stellar peaks under the cored halo tend to be sharper than under the cuspy halo. We also emphasize that these density profiles could be slightly altered by different initial conditions, which are randomly sampled from distribution functions using \texttt{AGAMA} as mentioned above. Furthermore, there are still density variations when the Indus progenitor is placed at the mid-point of $\phi_1$ $\sim$ $0\degr$. Finally, we show the case of OnlyStars in the bottom panel, which presents a fairly smooth stream with no major density fluctuations.

\subsection{Epicycles}

It is surprising to see that the models already exhibit similar episodic density variations to those of data given that we have not introduced any external perturbations. This implies that the profile might be a natural consequence from internal dynamics, specifically, the epicyclic motions \citep{2008MNRAS.387.1248K}. The tidal stripping process of stars may not be steady but highly fluctuating, especially for a dwarf galaxy with deep gravitational well in its centre. After evaporating from the progenitor, stars perform epicyclic motions due to pseudo-forces plus the tidal force, in which the stars will be periodically accelerated and decelerated while moving along the tidal tails \citep{2012MNRAS.420.2700K}. This leads to an oscillatory path which can be approximated by a smooth motion of a guiding centre about the Galactic centre superimposed by an oscillation about this guiding centre \citep{2008gady.book.....B}. The epicycles can result in density enhancements in streams, where overdensities correspond to the locations where the escaping stars slow down in their epicyclic motions, while underdensities happen where stars move fastest. 

\begin{figure*}
    \includegraphics[width=0.9\linewidth]{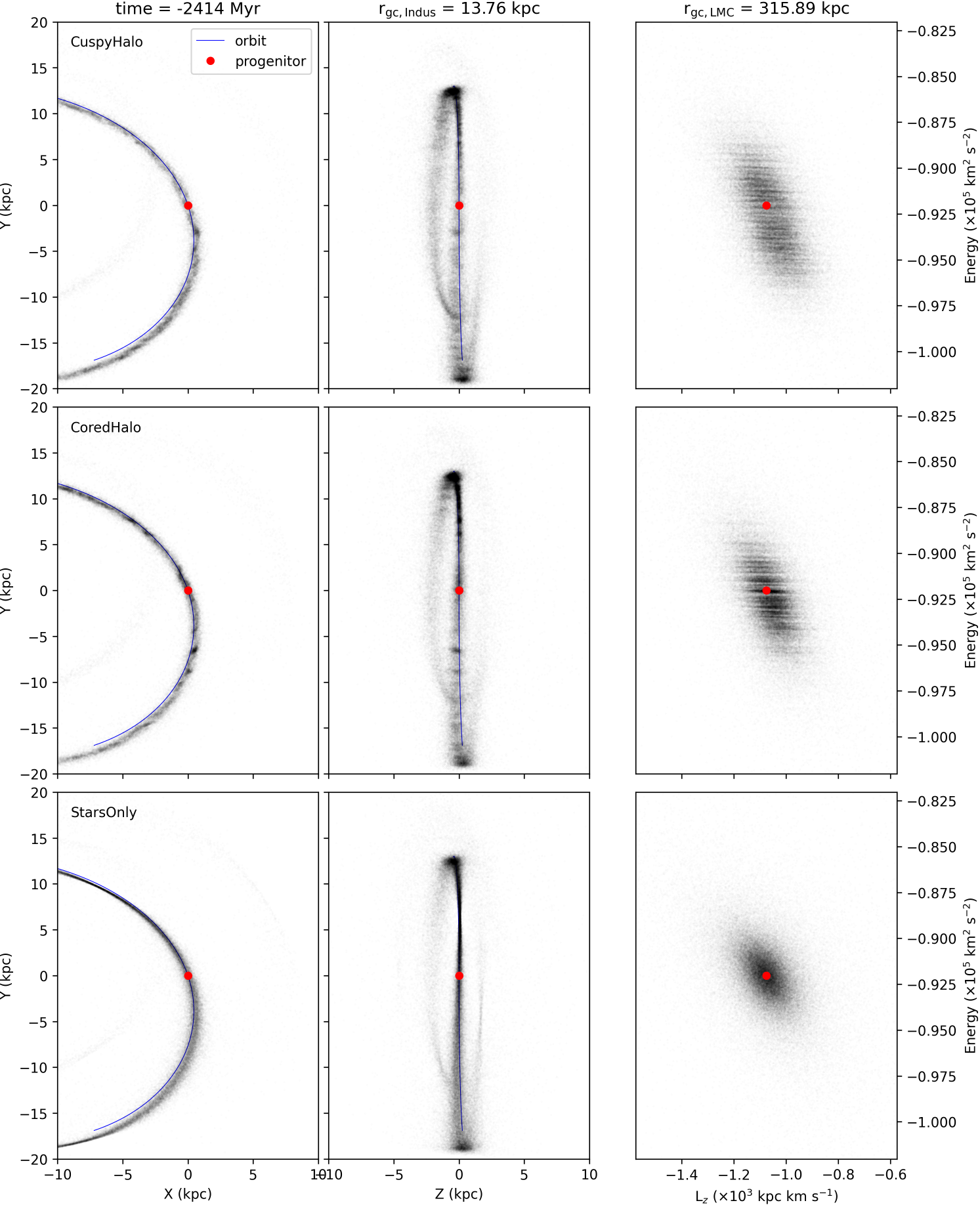}
    \caption{Distributions of the snapshots at midpoint of time for three modelling cases. Three rows correspond to snapshots of N-body simulations with the cuspy halo, the cored halo, and only stars, respectively. The first two columns show projections of (X, Y) and (Z, Y), where X and Y denote the instantaneous orbital plane of the progenitor, and Z is the normal. The third column shows the space of z-component angular momentum and energy in the external \texttt{potMWLMC} potential. Stellar particles of snapshots are shown in black dots, while the progenitor and its orbit are shown with red circles and blue lines. We further mark the time and Galactocentric distances of Indus and LMC on the top. In addition, we provide three corresponding animations to show the whole disruptions in supplementary materials, in which there will be three representative stars marked in green to show their epicyclic orbits.}
    \label{fig:epicycle}
\end{figure*}

To illustrate epicycles more directly, we extract snapshots that happen at the midpoint of time $-2.4$ Gyr for all three cases, as plotted in Figure~\ref{fig:epicycle}. We build a Cartesian coordinate always aligned with the stream, where X is from the Galactic centre to the stream progenitor but centred at the latter, Z is in the direction of total angular momentum, and Y is determined by $\mathbf{Z} \times \mathbf{X}$. Thus X and Y denote the instantaneous orbital plane of the progenitor and Z is the normal. In Figure~\ref{fig:epicycle}, the first two columns show projections of (X, Y) and (Z, Y), and the third column shows the space of integral of motions, namely z-component angular momentum and energy in the external \texttt{potMWLMC} potential. Three rows correspond to snapshots of N-body simulations for the case CuspyHalo, CoredHalo, and StarsOnly, respectively. Stellar particles are shown in black dots, along with the progenitor and its orbit shown with red circles and blue lines. We further mark the time and Galactocentric distances of Indus and LMC on the top.

By looking at the spatial distribution on the top row, it is easy to recognize a strong oscillatory pattern in the stream, showing a series of density variations. This is a signature of stars doing epicyclic motions along the tidal tails, and resembles the features pointed out in \citet{2012MNRAS.420.2700K}. In the integral of motion space, it is seen that a series of stripe-like structures appear along specific energy levels. Stars belonging to a certain energy stripe will oscillate as a group, and sometimes they may slow down together in epicycles to form density clumps. Compared to the snapshot for the CuspyHalo, the middle row for the CoredHalo presents some sharper density enhancements, in agreement with what is shown in Figure~\ref{fig:density_phi1}. There are also energy stripes which are more compact and enhanced. Turning to the bottom row for the case of StarsOnly, we do not observe a clear epicyclic feature, but this does not mean its absence. In supplementary materials, we provide three animations to illustrate the whole tidal disruptions. Besides all plots in Figure~\ref{fig:epicycle}, each animation further marks a representative star (green) to display its epicyclic orbit. The star will move following an oscillatory path with different amplitudes for three cases, which is clearer when observing in (Z,~Y) plane. The energy shock can be noted as well when close to the infall of LMC.

\subsection{Density Discrepancies}

\subsubsection{CuspyHalo vs. CoredHalo}

We start to compare the dwarf-like cases CuspyHalo and CoredHalo. The tidal radius is given by
\begin{equation}
r_t = \left( \frac{GM_\mathrm{prog.}}{4\Omega^2 - \kappa^2} \right)^\frac{1}{3} 
\end{equation}
where both $\Omega$ (angular velocity) and $\kappa$ (epicyclic frequency) are solely related to the progenitor orbit, meaning $r_t \propto M_\mathrm{prog.}^{1/3}$. Since stars are embedded in the inner region of the dwarf, different mass distributions at small radii between two dark matter halos will create different disrupting thresholds for stars. In detail, the increasing central density of the cuspy halo leads to more massive $M_\mathrm{prog.}$, in contrast to the flat inner density of the cored halo. It means that stars in case CuspyHalo have larger $r_t$ than in case CoredHalo. Stripped stars under the cuspy halo have overcome stronger internal gravity and experienced greater energy changes during the journey outward. This is why stream stars in case CuspyHalo span a wider range in energy as shown in Figure~\ref{fig:epicycle}. On the other hand, it is easier for stars in the cored halo to eject out due to smaller $r_t$, which causes more violent disruptions. For this reason, there are more stars contained within each energy stripe in Figure~\ref{fig:epicycle}, and thus some final density peaks comprise more stars to be sharper in Figure~\ref{fig:density_phi1}.

\begin{figure}
    \includegraphics[width=\columnwidth]{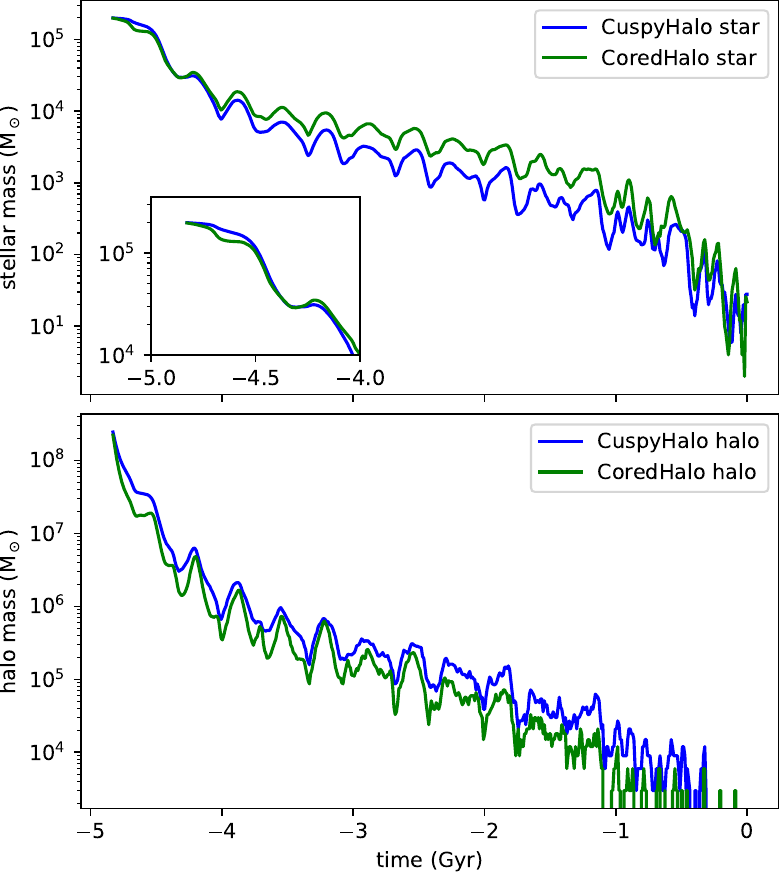}
    \caption{Evolutions of bound mass for CuspyHalo and CoredHalo. The upper panel shows stellar mass over time, where the subpanel is a zoom-in from $-5$ to $-4$ Gyr. Blue and green lines show the case CuspyHalo and CoredHalo, respectively. The lower panel is similar but for halo mass. }
    \label{fig:mass_loss}
\end{figure}

To quantify the mass-loss process, we calculate kinetic $+$ potential energy of all particles relative to the remaining progenitor and judge whether particles are bound with the relative total energy $<0$. Here potential between particles is calculated using \texttt{pytreegrav} \citep{2021JOSS....6.3675G}. Figure~\ref{fig:mass_loss} shows evolutions of bound stellar and halo mass for CuspyHalo and CoredHalo. In practice, we observe that both cases show very quick full dissolutions at around $-4$ Gyr, after which masses nearly stabilize. The time range only contains $\sim 3$ orbital periods (Indus period = 330 Myr). Because particles are orbiting the inner Milky Way between 12 kpc (pericentre) and 18 kpc (apocentre), the Galactic tidal field dominates over the dwarf internal structure. We expect the difference in lifetime between cuspy and cored halos to become significant at outer Galactocentric distances (e.g., $> 50$ kpc). There are periodic increases and decreases in mass due to stars contracting at apocentre and expanding at pericentre. It is seen that the cored halo mass goes down slightly faster than the cuspy halo mass because of their different central density profiles. Stars in CoredHalo are also stripped a little more quickly than in CuspyHalo before the full disruption at $\sim -4$ Gyr. After that, we are counting a specific segment of stars near the progenitor's orbit, where more stars of CoredHalo are there (similar to more stars within an energy stripe in Figure~\ref{fig:epicycle}). 

We can rule out a variable mass loss rate as the origin of the density clumps. The striped pattern in the energy visible in Figure~\ref{fig:epicycle} is mainly established between the third and fourth orbital periods. Three to four episodes of pericentre mass loss are not enough to produce the number of stripes in the energy distribution, and the stripes do not form at the same time as the pericentre passage, either. Moreover, there is long time of $>3$ Gyr for episodic stripping to keep mixing \citep{2012A&A...546L...7M,2021MNRAS.502.2364B} within the stream up to now. Therefore, we expect that mass loss is of minor importance in shaping final density variations of Indus. No matter what mass-loss rate is, stripped materials undergo mixing process along the stream by conducting epicyclic motions, and density clumps build up meanwhile.

\begin{figure}
    \includegraphics[width=\columnwidth]{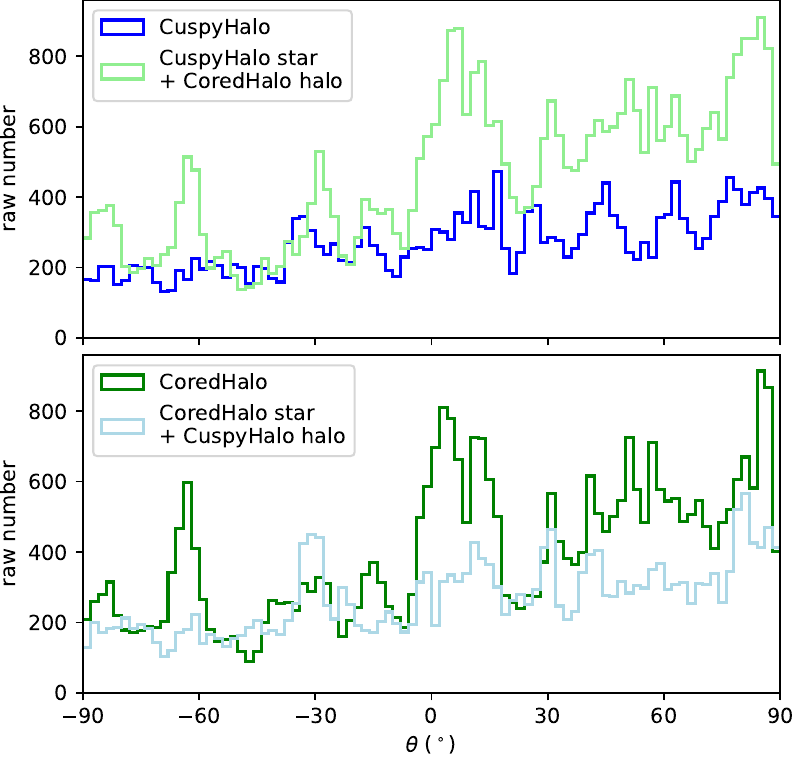}
    \caption{Density variations along streams as viewed from the Galactic centre. Case CuspyHalo and CoredHalo are simulations the same as in Figure~\ref{fig:density_phi1}. ``CuspyHalo star $+$ CoredHalo halo'' in the upper panel means that stars from case CuspyHalo evolve under the cored halo from CoredHalo. ``CoredHalo star $+$ CuspyHalo halo'' in the lower panel means stars from CoredHalo combined with the cuspy halo of CuspyHalo. These exchanges make sure that stellar components are the same and difference in stream densities is caused by the dark matter halo of the dwarf.}
    \label{fig:exchange}
\end{figure}

Distribution functions for stellar components, from which \texttt{AGAMA} draws stars, are slightly different because they are related to the total potentials of the dwarf system and different scale radii are used in Plummer spheres as well. To investigate that discrepancy in stream densities shown in Figure~\ref{fig:density_phi1} are truly caused by dark matter halos instead of stellar distributions, we exchange stellar components between cases CuspyHalo and CoredHalo, and rerun two simulations, and plot final results in Figure~\ref{fig:exchange}. Here $\theta$ denotes the angle in (X,~Y) plane between X and the line connecting from the Galactic centre to a star ($\theta$ continues to accumulate as the stream stretches longer). In the upper panel, ``CuspyHalo star $+$ CoredHalo halo'' means that stars from case CuspyHalo evolve under the cored halo from CoredHalo, such that stellar components are exactly the same. Similarly, ``CoredHalo star $+$ CuspyHalo halo'' in the lower panel means stars from CoredHalo combined with the cuspy halo of CuspyHalo. In general, we note that the final epicyclic density is dominated by the dark matter halo of the dwarf regardless of how stars look like. Overdensities generated under the cored halo always tend to contain more stars being sharper.

\begin{figure}
    \includegraphics[width=\columnwidth]{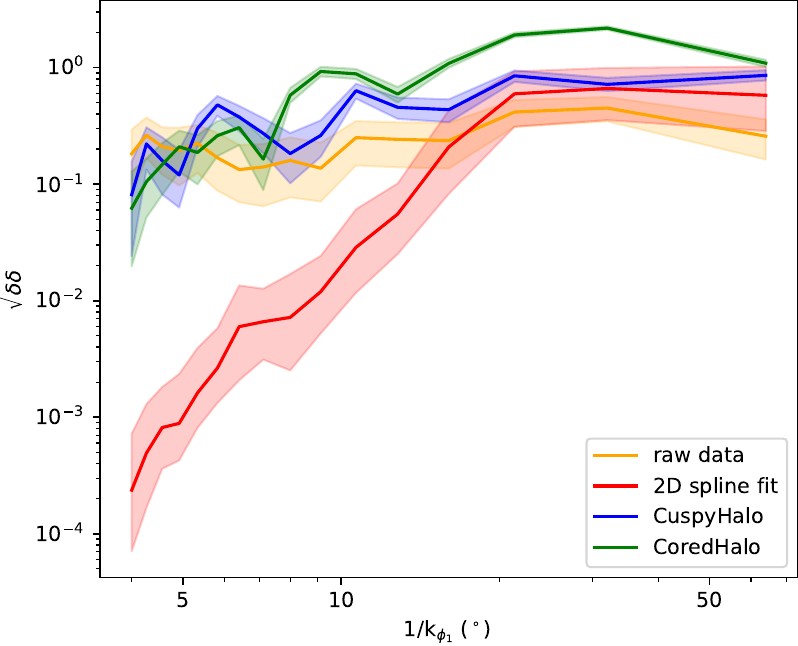}
    \caption{The power spectral densities as functions of inverse wavenumber $1/k_{\phi_1}$ for data and models. The orange and red come from the the histogram and spline fit of data, as shown in the first two panels of Figure~\ref{fig:density_model}. The blue and green correspond to CuspyHalo and CoredHalo of N-body results, as shown in the third and forth panels of Figure~\ref{fig:density_phi1}. All bands along with solid lines represent 16th, 50th, and 84th percentiles of 1000 samples.}
    \label{fig:power_spectra}
\end{figure}

Based on the density comparisons in Figure~\ref{fig:density_phi1}, we note that instead of sharp peaks caused by the cored halo, softened peaks from the cuspy halo are more favoured by data, implying that the Indus progenitor might have possessed a cuspy dark matter halo in the past. We can quantify how well they match one another by performing power spectrum analysis \citep{2017MNRAS.466..628B,2020ApJ...891..161I,2021MNRAS.502.2364B}. For data, we use density profiles shown in the first two panels of Figure~\ref{fig:density_model}, where the histogram serves to be a ``noisy'' version of the density from spline fit. For the histogram 1000 samples are generated by assuming Poisson uncertainty, and for the fit we randomly pick out 1000 realizations from its posterior distribution. These densities are normalized by dividing the third-order polynomial (black dashed lines in Figure~\ref{fig:density_model}). For two N-body models, we first apply proper-motion selections (ii) and (iii), and then sample 1000 times on their histograms against $\phi_1$ using Poisson uncertainty as well, and further apply the right term of Equation~\ref{eq:normalize} to samples, and finally divide them by the third-order polynomial (black dashed lines in Figure~\ref{fig:density_phi1}). We use \texttt{scipy.signal.welch} \citep{welch1161901} to compute the power spectral density. The 16th, 50th, and 84th percentiles of results as functions of inverse wavenumber $1/k_{\phi_1}$ are shown in Figure~\ref{fig:power_spectra}.

There are some random fluctuations at smaller angular scales around $5\degr$ for histograms of data and models, while the spline fit almost does not contain them due to its smooth trends. A slightly stronger power at $\sim 10\degr$ may be due to some peaks at that scale, for example, the sharp peak at $\phi_1 = -20\degr$ from CoredHalo. Major density variations from epicycles typically span $\sim 20\degr$, and a rising power is revealed at that scale for both data and models. Importantly, it is seen that the data spectra are more consistent with CuspyHalo rather than CoredHalo.

\subsubsection{StarsOnly}

We then take a look at the case StarsOnly. According to \citet{2010MNRAS.401..105K}, the separation between epicyclic overdensities is
\begin{equation}
y_c = \frac{4\pi \Omega}{\kappa} \left( 1-\frac{4\Omega^2}{\kappa^2} \right) r_t 
\propto M_\mathrm{prog.}^\frac{1}{3}  .
\end{equation}
As a result, the epicyclic amplitude is much smaller for only stars in the absence of a massive dark matter halo. Due to strong tidal field outside and diffuse stellar distributions inside (scale radius = 0.75 kpc), the Indus core is immediately dissolved and is stretched longer slowly. In other words, initial stars at the beginning are almost unbound given that we do not observe their energy changes due to disruption, which makes this model sort of unphysical. Although stars perform epicyclic motions individually, they occupy spread and mixed orbits, and the resultant distribution of $y_c$ (instead of a constant $y_c$) just smooths out the longitudinal density. Therefore, epicycles become unpronounced in Figure~\ref{fig:density_phi1} and Figure~\ref{fig:epicycle}, but are still subtly present in the supplementary animation.

\subsubsection{Velocity Dispersion}

\begin{figure}
    \includegraphics[width=\columnwidth]{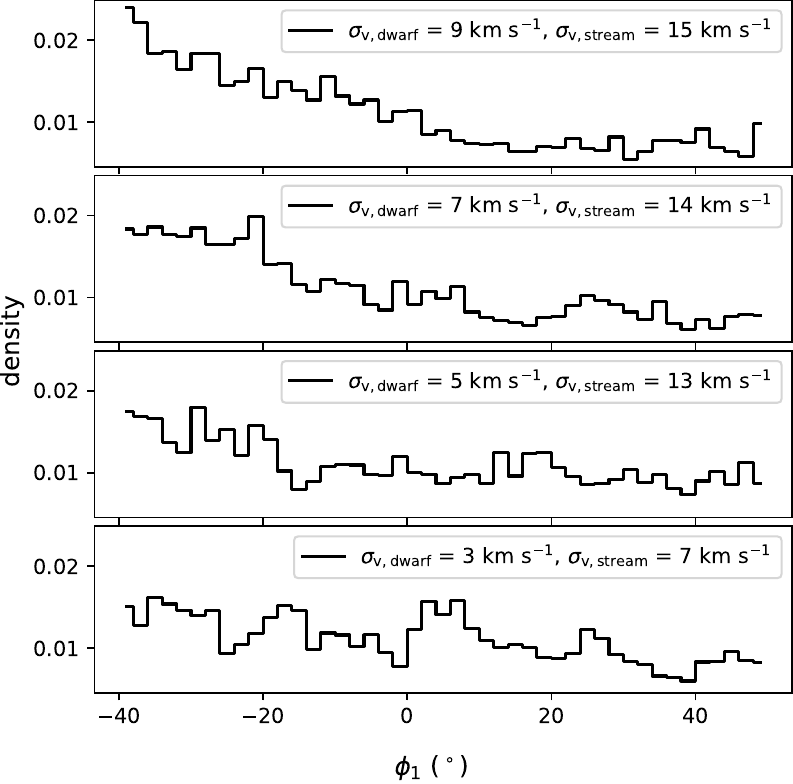}
    \caption{Stream densities of multiple Indus-like simulations with increasing velocity dispersions. The bottom one corresponds to CuspyHalo, and the above three are similar to CuspyHalo but using different mass and scale radius for halo and stellar components (see text for details).}
    \label{fig:test4vdisp}
\end{figure}

Epicyclic density variation theory is established and focused on globular cluster tidal tails \citep[e.g.,][]{2008MNRAS.387.1248K,2012A&A...546L...7M,2017MNRAS.470...60E,2020ApJ...891..161I}. Here we apply it to dwarf galaxy streams. Mass loss is achieved by different mechanisms, stars evaporating through two-body relaxation for globular clusters, while tidal stretching for dwarf galaxies \citep{2012MNRAS.420.2700K}. However, stripped stars no longer have exactly the same orbits as the progenitor, such that we always expect to observe their oscillatory motions in the reference frame comoving with the progenitor. Thus, the epicycle should be common, and the question would be when it matters in stream densities. In \citet{2010MNRAS.401..105K} and \citet{2012MNRAS.420.2700K}, it is pointed out that the key to make epicyclic clumps vanish is the scatter in escape conditions of escaping stars. Based on this point we test another three simulations similar to CuspyHalo with increasing velocity dispersions. Specifically, halo and stellar mass are ($3\times10^8$, $2\times10^5$), ($6\times10^8$, $5\times10^5$), and ($9\times10^8$, $7\times10^5$) M$_{\odot}$ according to the stellar-halo-mass relation from \citet{2017MNRAS.467.2019R}, with halo and stellar scale radius (1.0, 0.5) kpc for all of them. These settings give initial stellar velocity dispersions of 5, 7, and 9 km~s$^{-1}$, along with present-day stream dispersions of 13, 14, and 15 km~s$^{-1}$. As a reference, CuspyHalo has a initial 3 and present-day 7 km~s$^{-1}$, respectively. Their raw densities along $\phi_1$ are compared in Figure~\ref{fig:test4vdisp}. It is noted that these variations get smoother as velocity dispersions go higher, and density clumps may be eventually washed out when a dwarf galaxy stream owns a dispersion of, e.g., $>15$ km~s$^{-1}$.

\subsection{External Effects}

Based on our modelling, we find that natural evolution already creates density fluctuations comparable to the measurements from the data, that is, there are the same numbers of peaks and gaps, and their locations approximately match as well. Therefore, we consider that the internal epicycles might be the main reason. However, this does not completely exclude external perturbations such like dark matter subhalos, and we discuss underlying signatures for them in the following.

\begin{figure}
    \includegraphics[width=\columnwidth]{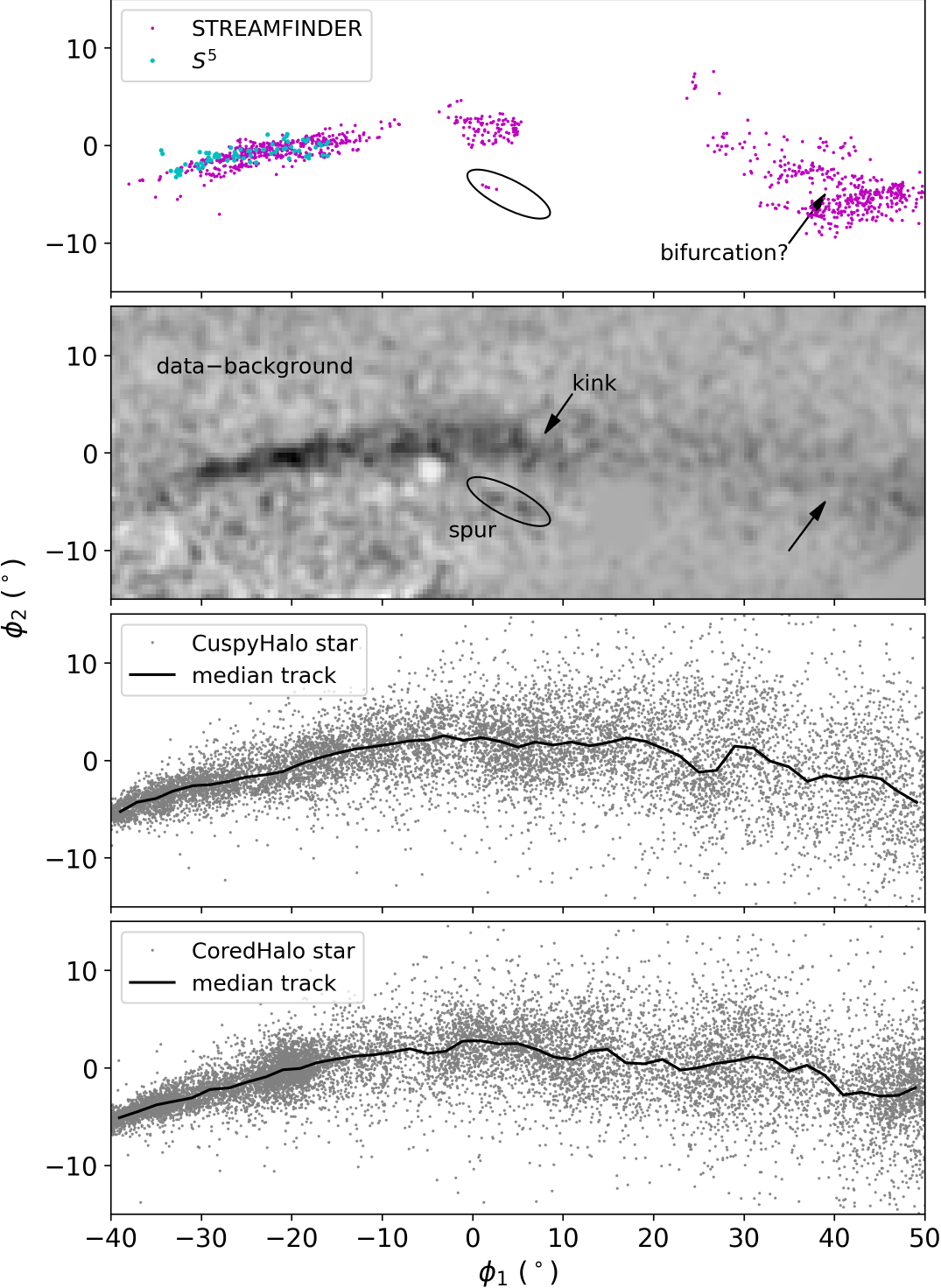}
    \caption{Comparisons of Indus morphology in data and models. Indus members from $S^5$ \citep{2022ApJ...928...30L} and \texttt{STREAMFINDER} \citep{2024ApJ...967...89I} are shown in the first panel. The second panel shows the filtered map after subtraction of fitted background, the same as in Figure~\ref{fig:density_comparison}. The ellipse and arrows indicate some possible smaller-scale features of Indus. Model streams of CuspyHalo and CoredHalo are plotted in grey dots in last two panels, where balck lines are their median tracks (medians are calculated in $2\degr$ wide $\phi_1$ bins).}
    \label{fig:if_external}
\end{figure}

Figure~\ref{fig:if_external} compares morphologies of Indus in data and our models. Indus members from $S^5$ \citep{2022ApJ...928...30L} and \texttt{STREAMFINDER} \citep{2024ApJ...967...89I} are shown in the first map, where we see that spectroscopic targets of $S^5$ are confined to the most negative $\phi_1$ end of the DES coverage, while \texttt{STREAMFINDER} stars consist of three clumps which could be separated by SMC and LMC. The second panel shows the filtered map after subtraction of the fitted background, the same as in Figure~\ref{fig:density_comparison}. Indus streams of CuspyHalo and CoredHalo are plotted in grey dots in last two panels, where black lines indicate their median tracks (medians are calculated in $2\degr$ wide $\phi_1$ bins). Firstly, it is worth noting that there is a small overdense region marked by the black ellipse, where five stars from \texttt{STREAMFINDER} are located. The signal appears in every spatial filtered map and may belong to Indus as a part outside the main stream body (often called spur), probably a result of a subhalo flyby. Secondly, a kink seems to appear within the stream near $\phi_1=10\degr$, and we note that the trajectory measurement also indicates a slight down-turn at the same location in Figure~\ref{fig:density_model} ($\phi_2$ panel). The feature is right above the spur, which might be perturbed by the same flyby. However, the kink can be attributed to epicycles as well, because there are similar twists recovered by N-body models as shown by black median tracks in lower two panels. Thirdly, \texttt{STREAMFINDER} stars seemingly indicate a bifurcation at $\phi_1=40\degr$, but we can not confirm it so far. Stars are incomplete at $\phi_1<35\degr$, and the feature is not clear in our data, either. It might be just due to stars being too sparse given that $\phi_1=40\degr$ corresponds to an underdensity.

Additionally, we conduct a few experiments to inspect effects from baryonic structures in the Milky Way. Briefly, we use an isotropic Gaussian ball of particles \citep{2017NatAs...1..633P,2022ApJ...941..129D} to generate a simplified model stream owning a flat longitudinal density, making it easy to recognize any perturbations. Particles are produced by sampling positions and velocities from three dimensional Gaussians with dispersion = 0.1 kpc and 7 km~s$^{-1}$, respectively. They are placed on the best-fit Indus orbit to evolve from $-5$ Gyr up to now under the \texttt{potMWLMC} without mutual gravity considered. Although this model is similar to a globular cluster, we expect that the final stream can roughly mimic the overall motion, width and velocity dispersion of Indus. We follow methods used in \citet{2021ApJ...911..149L} to incorporate gravitational influences from baryons. The Galactic bar and spiral arms can be included in the Milky Way potential from \citet{2024A&A...692A.216H}. Giant molecular clouds are fetched from \citet{2017ApJ...834...57M} and distributed in the Galactic disk. Dwarf galaxies are taken from \citet{2025OJAp....8E.142P}, while globular clusters are from \citet{2021MNRAS.505.5978V} and \citet{2021MNRAS.505.5957B}. For the latter three types, we exclude those whose orbits never fall between Galactocentric distances of 10 kpc and 20 kpc, given Indus's pericentre = 12 kpc and apocentre = 18 kpc. We repeat our modelling of Indus in the presence of these respective components individually, starting from $-500$ Myr to now and covering one and a half orbital periods of Indus ($\simeq$ 330 Myr). Overall, we find fairly flat density profiles quite close to the unperturbed one, with none of above baryonic components able to cause large density changes. This is foreseeable because: for the bar, spiral arms, and molecular clouds confined in the disk, Indus is orbiting far from the Galactic centre with a pericentre = 12 kpc and is highly inclined with an orbital pole pointing to $b=-20\degr$; the classic dSph Sagittarius mainly dominates the halo beyond 20 kpc that is farther than Indus apocentre = 18 kpc, and the other dwarfs do not encounter the stream closely, either; globular clusters are not massive enough to produce significant features in the dynamically ``hot'' Indus.

\section{Future Prospects} \label{sec:future}

\subsection{Modelling Refinement}

In Figure~\ref{fig:density_phi1} we claim a qualitative consistency between data and models, that is, there is still somewhat mismatch for the central locations of peaks and gaps. During the process of optimising N-body models, we note two important factors that can affect final longitudinal densities.

\begin{enumerate}

 \item The stream orbit matters. How the system moves definitely determines how it experiences the tidal force from the Milky Way, especially for \texttt{potMWLMC} that is time-dependent. The Indus dwarf is supposed to fall inward gradually from the very outskirt of the halo due to dynamic friction, which is not considered in our models. More realistic modelling in the future needs more precise orbit determination by further taking into account the dynamic friction.

 \item Initial conditions matter. We also find that final densities are closely related to positions and velocities of particles (especially dark matter) in the initial dwarf system. Due to limited resolution (e.g., $3\times10^3$ M$_{\odot}$ for dark matter), particles in the phase space can slightly differ from time to time because they are sampled with certain distribution functions. As a result, the final density status can be adjusted accordingly. Here distribution functions are directly from mass profiles of Indus, regulated by which mass model to use along with its parameters. Improvements in initial conditions can help to refine the mismatch between data and models.

\end{enumerate}

\subsection{North Indus}

\begin{figure}
    \includegraphics[width=\columnwidth]{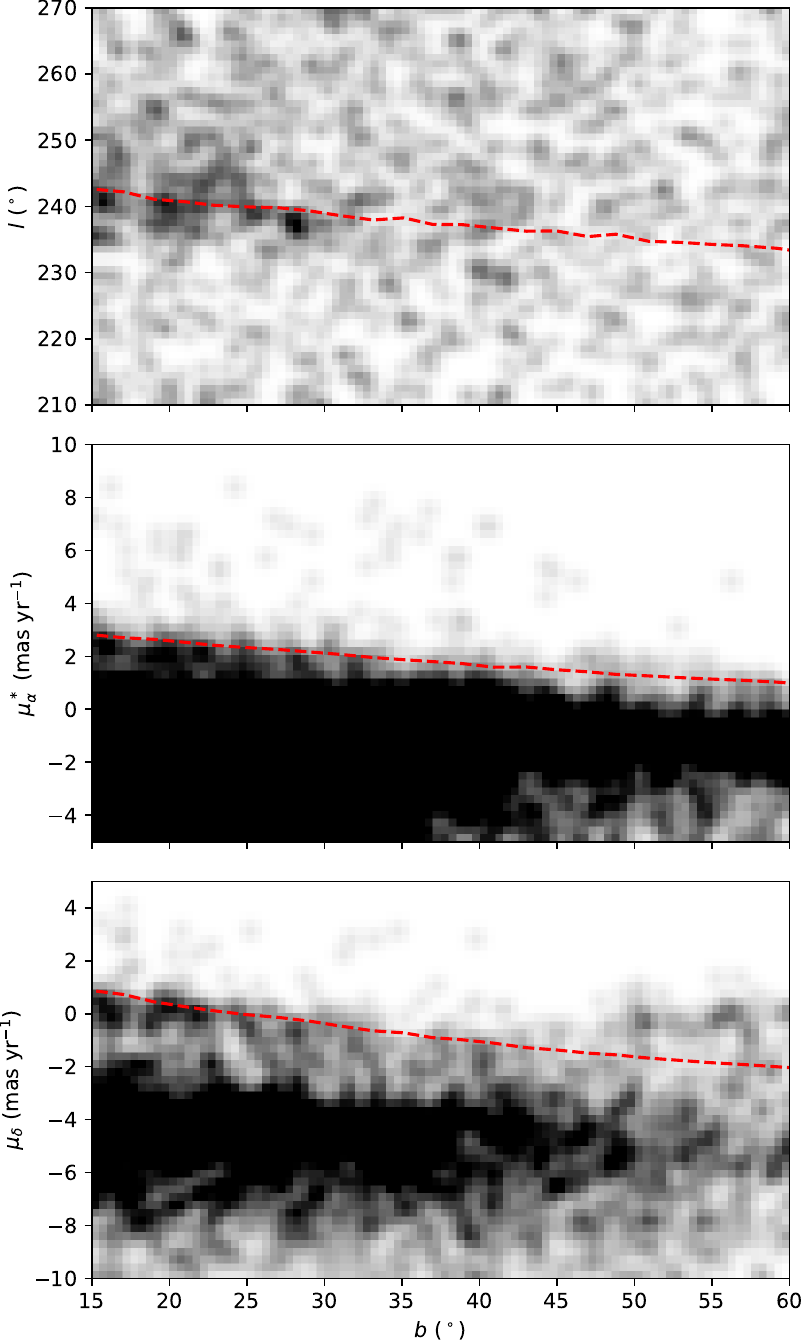}
    \caption{Filtered maps in $l$, $\mu^*_\alpha$, and $\mu_\delta$ as a function of $b$, similar to Figure~\ref{fig:indus_kinematic_cmd} but for the north Indus. The red lines are predictions based on the trial model.}
    \label{fig:northIndus}
\end{figure}

We have focused on Indus in the southern Galactic sky, but the stream does not necessarily end at the point that crosses the Galactic plane. We therefore have carried out a brief search for Indus in the northern Galactic sky. We retrieve \textit{Gaia} data above Galactic latitude $b>15\degr$ overlapping with the trial model stream. We then process the data following the steps in Section~\ref{sec:data} and Section~\ref{subsec:filtering}. To show its extent, we plot filtered maps in Figure~\ref{fig:northIndus}, similar to Figure~\ref{fig:indus_kinematic_cmd}. Each dimension is the result after filtering in other dimensions plus in colour-magnitude space. We use $b$ as the longitude that is almost parallel to the stream. The red dashed lines indicate stream tracks predicted by the trial model.

We find that Indus does likely extend across the Galactic plane. There are overdense pixels around the tracks near the disk side. The spatial one is quite diffused, which is the same behaviour as the south part when close to the disk. From $\mu^*_\alpha$ we notice a trend gradually falling down to contaminations as $b>33\degr$. Fortunately, $\mu_\delta$ can presents results more clearly, and it is seen that the stream may continue until $b\sim40\degr$. We do not perform morphology fitting due to its scattered distribution and short extent. However, we can recognize by eye that there are also density variations here, for example an underdensity at $b \sim 35\degr$.

\section{Conclusion} \label{sec:summary}

In this paper, we focus on the Indus stellar stream, the tidal remnant of a former dwarf galaxy in the Milky Way halo. Our main findings are summarised in the following.

\begin{enumerate}

 \item Taking advantage of \textit{Gaia} DR3 and with the help of $S^5$ data, we perform a comprehensive matched-filter analysis to detect the Indus stream. The filtering is applied in celestial, proper-motion, and CMD spaces, by which a $\sim 90\degr$ Indus is revealed in the southern Galactic sky. The length is improved compared to original $\sim 20\degr$ from \citet{2018ApJ...862..114S} and in concordance with the extent from \citet{2024ApJ...967...89I}. The structure terminates as it approaches the Galactic bulge, whereas the opposite end goes across the Galactic disk. We further present a preliminary matched-filter detection of the north Indus and show evidence of the stream extending towards galactic latitude $b\sim40\degr$.
 
 \item To extract the stream's morphology, a density model consisting of a stream and a background is fitted to the spatial filtered map. The resulting model successfully reproduces the principal features observed in the data. The fit quantitatively presents the density, trajectory, and width varying along the stream. Strong density variations are observed with episodic peaks and gaps in Indus.
 
 \item We model the Indus stream using N-body simulations. The initial conditions contain both stellar and dark matter particles as Indus has a dwarf-galaxy origin. Stellar mass is estimated to be $\sim 2 \times 10^5$ M$_{\sun}$ which roughly corresponds to a dark matter halo mass of $\sim 3 \times 10^8$ M$_{\sun}$. After tracking tidal disruptions for nearly 5 Gyr, we find the model streams to exhibit density fluctuations comparable to the data measurement, with the same numbers and similar locations of peaks and gaps. We observe evident epicyclic motions of stars happening during the stream stretching longer, which account for the final longitudinal density variation.
 
 \item In N-body simulations, we assign stars either a cuspy or a cored dark matter halo by setting the inner slope $\gamma$ to be 1 or 0 in Equation~\ref{eq:nfw}. Epicyclic density clumps in the cuspy case appear to be mild, compared to likely sharper peaks in the cored case. This is because compared to the cuspy halo, stars in the cored halo are less tightly bound, which causes more drastic instantaneous mass loss. Such mass loss makes stream overdensities contain more stars to be sharper. We use power spectra to quantify how well densities from data match with models. The cuspy result is found to be more consistent with data, implying that Indus might have possessed a cuspy halo before disruption.

 \item The theory of epicyclic density variations is originally developed for globular cluster tidal tails. Here we apply it to dwarf galaxy streams. Although both kinds have epicycles, the latter usually own higher velocity dispersions, being less sensitive to external perturbations. Therefore, instead of dwarf galaxy streams, cold globular cluster streams might be more ideal tools to probe dark matter subhalos.

\end{enumerate}

We further attach three animations in supplementary materials to show epicyclic motions in the three types of N-body simulations. Overall, in-depth investigations into longitudinal densities along dwarf galaxy streams remain an area with significant additional room for exploration. Serving as an initial attempt to prompt deeper inquiry, we focus on the density profile of Indus in this work. We anticipate that deeper and higher-quality data in the near future would help to confirm the irregular density fluctuations of Indus, for instance, the discontinuity shown at $\phi_1 = -27\degr$ in Figure~\ref{fig:gap_G21mag}. In addition, more details about the morphology of Indus may be revealed, such as the spur, kink, or possible bifurcation in Figure~\ref{fig:if_external}. These will help us understand the dwarf galaxy debris populating the inner Milky Way.

\section*{Acknowledgements}

We thank the anonymous referee for constructive suggestions that greatly improve this work. Y.Y. thanks Jordan Winstanley and Smrithi Gireesh Babu at the University of Sydney for kind discussions on \texttt{GADGET-4} settings. Y.Y., G.F.L., S.L.M., and D.B.Z. acknowledge support from the Australian Research Council through the Discovery Program grant DP220102254. S.L.M. acknowledges the support of the UNSW Scientia Fellowship Program.

This paper includes data obtained with the Anglo-Australian Telescope in Australia. We acknowledge the traditional owners of the land on which the AAT stands, the Gamilaraay people, and pay our respects to elders past and present.

This work presents results from the European Space Agency (ESA) space mission \textit{Gaia}. \textit{Gaia} data are being processed by the \textit{Gaia} Data Processing and Analysis Consortium (DPAC). Funding for the DPAC is provided by national institutions, in particular the institutions participating in the \textit{Gaia} MultiLateral Agreement (MLA). The \textit{Gaia} mission website is \url{https://www.cosmos.esa.int/gaia}. The \textit{Gaia} archive website is \url{https://archives.esac.esa.int/gaia}.

\section*{Data Availability}

This paper mainly relies on publicly available \textit{Gaia} DR3. The Indus members of $S^5$ and \texttt{STREAMFINDER} come from \citet{2022ApJ...928...30L} and \citet{2024ApJ...967...89I}, respectively.



\bibliographystyle{mnras}
\bibliography{indus} 



\bsp	
\label{lastpage}
\end{document}